\documentclass[aps,preprint,epsfig,rotate]{revtex4}
\usepackage{graphicx}
\usepackage{bm}
\usepackage{epsfig}

\input epsf
\begin{document}
\title{Properties of the weakly-bound (1,1)-states and rotationally excited (2,0)-states in the muonic molecular 
       $d d \mu, d t \mu$ and $t t \mu$ ions}

\author{Alexei M. Frolov}
 \email[E--mail address: ]{alex1975frol@gmail.com} 


\affiliation{Department of Applied Mathematics \\
 University of Western Ontario, London, Ontario N6H 5B7, Canada}

\date{July 27, 2023}

\begin{abstract}

Total energies and other bound state properties of the weakly-bound (1,1)-states and rotationally 
excited (2,0)-states in the three-body muonic molecular $d d \mu, d t \mu$ and $t t \mu$ ions are 
determined to high numerical accuracy and investigated. Our current numerical accuracy achieved 
for the total and binding energies of the weakly-bound (1,1)-states in the both $d d \mu$ and 
$d t \mu$ ions significantly exceeds similar accuracy obtained in earlier computations of these 
weakly-bound states in these two ions. The bound state properties of the weakly-bound (1,1)-states 
and (2,0)-states in the $d d \mu, d t \mu$ and $t t \mu$ muonic ions have never been determined to 
high accuracy in earlier studies. We also briefly discuss the current status of muon-catalyzed  
nuclear fusion and develop the new universal variational expansion which can be used for extremely 
accurate bound state calculations of arbitrary three-body systems, including the truly adiabatic 
${}^{\infty}$H$^{+}_{2}$ ion and close systems.


\end{abstract}

\maketitle


\section{Introduction}

Accurate and highly accurate calculations of bound state properties of various three-, few- and many-body systems 
always have a great interest in a large number of applications, since data from such computations can directly be 
compared with the known experimental results. In this study we restrict ourselves to the analysis of bound state 
properties of some muonic molecular ions each of which includes three electrically charged particles. Two of these 
particles are the heavy nuclei of hydrogen isotopes, i.e., protium $p^{+}$, deuterium $d^{+}$ and/or tritium 
$t^{+}$, while the third particle in each of these ions is the negatively charged muon $\mu^{-}$. In the 
lowest-order approximation each of these three particles can be considered as a point, structureless and 
non-relativistic particle. In this approximation there are six muonic molecular ions $p p \mu, p d \mu, p t \mu, 
d d \mu, d t \mu, t t \mu$ and 22 bound (stable) states in them (see, e.g., Table 6 in \cite{FroW2011}). In fact, 
each of the protium muonic molecular ions, i.e., the $p p \mu, p d \mu$ and $p t \mu$ ions, has two bound states: 
one $S(L = 0)-$state and one $P(L = 1)-$state. Similar two-state bound spectra are called the protium-type spectra 
of the muonic molecular ions. The bound state spectrum of each of the deuterium $d d \mu$ and $d t \mu$ ions 
includes five states: two $S(L = 0)-$states, two $P(L = 1)-$states and one $D(L = 2)-$state (deuterium-type spectra). 
The `vibrationally' excited $P(L = 1)-$state, which is designated as the $P^{*}(L = 1)-$state, or (1,1)-state, in 
these series of bound states is always weakly-bound \cite{Fro1992}. The heaviest muonic $t t \mu$ ion has six 
bound states: two $S(L = 0)-$states, two $P(L = 1)-$states, one $D(L = 2)-$state and one $F(L = 3)-$state 
(tritium-type spectra of the muonic molecular ions). None of these six bound states is weakly-bound. The existence 
of three different types of bound state spectra in three-body muonic molecular ions follows from the known fact, 
that the type of bound state spectrum in an arbitrary three-body $a b \mu$ ions is mainly determined by the 
lightest mass of the two heavy particles $a$ and $b$ \cite{Fro1992}, e.g., by the particle $a$, if $m_a \le m_b$.  
  
In many cases highly accurate computations of the total and/or binding energies and bound state properties in 
muonic molecular ions are difficult to perform. For instance, the two well-known weakly-bound (1,1)-states in the 
muonic $d d \mu$ and $d t \mu$ ions have never been determined to very high numerical accuracy, which currently 
for muonic molecular ions can be evaluated as $\approx 1 \cdot 10^{-22} - 1 \cdot 10^{-23}$ $m.a.u.$ (and higher), 
where the brief notation $m.a.u.$ means the muon-atomic units where $\hbar = 1, e = 1$ and $m_{\mu} = 1$ (also $4 
\pi \epsilon_0 = 1$). Such an accuracy has been achieved, e.g., for the ground states in muonic molecular ions, 
including the $S(L = 0)-$state(s) in the $p p \mu, d d \mu$ and $t t \mu$ ions and for some $P(L = 1)-$states. 
For weakly-bound (1,1)-states (or $P^{*}(L = 1)-$states) in the $d d \mu$ and $d t \mu$ ions the current 
(maximal) numerical accuracy can be evaluated as $\approx 1 \cdot 10^{-15}$ $m.a.u.$ and $\approx 1 \cdot 
10^{-13}$ $m.a.u.$, respectively \cite{FroPLA}. This is mainly related to slow convergence of the traditional 
variational expansions in applications to extremely weakly-bound states.   

Our main goal in this study is to perform highly accurate computations of the weakly-bound (1,1)-states in the 
muonic $d d \mu$ and $d t \mu$ ions and (2,0)-states (or $D(L = 2)-$states) in the heavy muonic $d d \mu, d t 
\mu$ and $t t \mu$ ions. Briefly, we want to improve (substantially) the overall accuracy achieved in previous 
calculations of these ions and determine (to high accuracy) a number of bound state properties for each of 
these ions. Another aim is to determine a number of bound state properties of the weakly-bound (1,1)-states in
the both muonic $d d \mu$ and $d t \mu$ ions and analogous properties for the bound (2,0)-states of the  
muonic $d d \mu, d t \mu$ and $t t \mu$ ions. Note that each of the muonic $a a \mu$ and $a b \mu$ ions is the 
Coulomb three-body system with unit electrical charges. Each of these systems has a finite number of bound 
states and the total number of bound states increases when the masses of two heavy positively charged particles 
increase. This follows from the general principles of bound states in the Coulomb three-body systems with unit 
charges \cite{Fro1992}. 

To designate the bound states in three-body muonic molecular ions $a b \mu$ in this study we shall apply the 
two-center (or molecular) system of notations $(\ell, \nu)-$notation, where $\ell (= L)$ is the rotational 
quantum number of this bound state, while $\nu$ is the 'vibrational' quantum number of the same state. 
Classification of bound states in terms of the `rotational' $\ell$ and `vibrational' $\nu$ quantum numbers 
allows one to designate (correctly and uniquely) an arbitrary bound state in all three-body muonic molecular 
ions $a a \mu$ and $a b \mu$, where $a$ and $b$ are the nuclei of hydrogen isotopes $p, d$ and/or $t$. 
Furthermore, such a system of $(\ell, \nu)-$notation has a number of advantages in applications to the heavy 
muonic molecular ions $a b \mu$ which can be considered (in the lowest-order approximation) as the two-center 
three-body systems. In order to illustrate this fact let us note that the weakly-bound (1,1)-states in the 
three-body $d d \mu$ and $d t \mu$ muonic ions in the one-center (or atomic) system of notations must be 
labeled as the excited $P^{*}(L = 1)-$states. The last notation is more complex and formally needs some 
additional explanations. 

The weakly-bound (1,1)-states in the both $d d \mu$ and $d t \mu$ muonic molecular ions play a central role 
in the `resonance' muon-catalyzed fusion (see, e.g., \cite{MS} and references therein). It can be illustrated 
by the fact that one negatively charged muon $\mu^{-}$ produces (or catalyzes) up to $\kappa \approx$ 150 - 
180 nuclear $(d,t)$-fusion reactions in the liquid, equimolar (or one-to-one, or 1:1) deuterium-tritium 
mixture. In liquid deuterium such a number of fusion reactions per one muon is smaller $\kappa \le 30$, but 
it is still significantly larger (in dozens of times) than experimentally known $\kappa$ number(s) for 
analogous $p d \mu$ and $p t \mu$ muonic ions where only non-resonance muon-catalyzed fusion can proceed (see, 
e.g., \cite{Alv}). The resonance muon-catalyzed fusion is briefly discussed in the Appendix A. Currently, it 
is absolutely clear that muon catalyzed nuclear fusion cannot be used for the energy production. However, 
theoretical and experimental investigations of the bound state properties and processes and reactions in all 
three-body muonic molecular ions and analogous four-, five- and six-body systems \cite{Fro56body} can supply 
us with a large volume of valuable information. For instance, by performing experiments with muonic molecular 
ions one can accurately evaluate the cross section of some important nuclear reactions at small and very small 
energies ($E \le$ 0.35 $keV$). In general, it is very difficult to measure such cross sections in the direct 
scattering experiments. Also, if one needs a compact, very intense and reliable source of thermonuclear 14.1 
$MeV$ neutrons, then the muon-catalyzed fusion in the liquid, equimolar DT-mixture is, probably, the best 
choice.  

In the Appendix B we consider some problems which are crucial in order to develop the actual universal expansions 
applicable to arbitrary three-body systems, including pure adiabatic molecular ions. This problem has been 
investigated in a number of earlier studies, but in this Appendix we are taking the final step on that long road. 
We have developed the new variational expansion(s) in the three-body relative and/or perimetric coordinates which 
is `absolutely universal' expansion for three-body systems. This expansion can be used for highly accurate, bound 
state calculations of all three-body ions, atoms and clusters. It works equally well and effective for three-body 
systems, including the pure adiabatic ${}^{\infty}$H$^{+}_{2}$ ion, `regular' Ps$^{-}$ and H$^{-}$ ions, He-like 
atoms/ions, muonic molecular ions $a b \mu$ and for many dozens (even hundreds) similar systems.    
     
\section{Hamiltonian and Exponential Variational Expansion}

The Hamiltonian of the three-body muonic $a b \mu$ ion is written in the form
\begin{eqnarray}
 \hat{H} = -\frac{\hbar^2}{2 m_{a}} \Delta_{a} - \frac{\hbar^2}{2 m_{b}} \Delta_{b} 
 -\frac{\hbar^2}{2 m_{\mu}} \Delta_{\mu} + \frac{e^2}{r_{ab}} - \frac{e^2}{r_{a\mu}} 
 - \frac{e^2}{r_{b\mu}} \; , \; \label{Hamilt1}
\end{eqnarray}
where $\Delta_{i} = \frac{\partial^{2}}{\partial x^{2}_{i}} + \frac{\partial^{2}}{\partial y^{2}_{i}} + 
\frac{\partial^{2}}{\partial z^{2}_{i}}$ is the Laplace operator of the particle $i$, while $\frac{e^2}{r_{ij}} 
= \frac{e^2}{\mid {\bf r}_{i} - {\bf r}_{i} \mid}$ is the Coulomb interaction between two point particles ($i$ 
and $j$). Also in this equation $\hbar = \frac{h}{2 \pi}$ is the reduced Planck constant, which is also called 
the Dirac constant, $e$ is the electric charge of the electron, $m_{\mu}$ is the rest mass of the negatively 
charged $\mu^{-}$ muon, while $m_a$ and $m_b$ are the nuclear masses of the two hydrogen isotopes. Below, we 
shall always assume that $m_a \le m_b$ and designate the muon $\mu^{-}$ as the particle 3, while the two 
positively charged particles $a$ and $b$ will be denoted as the particles 1 and 2, respectively. In muon-atomic 
units the same Hamiltonian $\hat{H}$ takes the form 
\begin{eqnarray}
 \hat{H} = -\frac{1}{2 X} \Delta_{1} -\frac{1}{2 Y} \Delta_{2} - \frac{1}{2} \Delta_{3} 
  - \frac{1}{r_{32}} - \frac{1}{r_{31}} + \frac{1}{r_{21}} \; , \; \label{Hamilt}
\end{eqnarray} 
where $X = \frac{m_a}{m_{\mu}}$ and $Y = \frac{m_b}{m_{\mu}}$ are the nuclear masses of hydrogen isotopes 
expressed in terms of the muon mass $m_{\mu}$.  

In this study all highly accurate computations of the weakly-bound (1,1)-states in the $d d \mu$ and $d t 
\mu$ muonic molecular ions and bound (2,0)-states in the heavy $d d \mu, d t \mu$ and $t t \mu$ muonic 
molecular ions are performed with the use of our exponential variational expansion in the relative and/or 
perimetric coordinates. The exponential variational expansion in the relative coordinates $r_{32}, r_{31}$ 
and $r_{21}$ has the following general form (see, e.g., \cite{Fro2001} and references therein)
\begin{eqnarray}
 \Psi_{LM} = \frac{1}{2} (1 + \kappa \hat{P}_{12}) \sum_{n=1}^{N} 
 \sum_{\ell_{1}} C_{n} {\cal Y}_{LM}^{\ell_{1} \ell_{2}}({\bf r}_{31}, {\bf r}_{32}) 
 \exp(-\alpha_{n} r_{32} - \beta_{n} r_{31} - \gamma_{n} r_{21}) \; \; , \; \label{exprel} 
\end{eqnarray}
where ${\bf r}_{ij} = {\bf r}_{i} - {\bf r}_{j} = - {\bf r}_{ji}, {\bf r}_{ji} = {\bf r}_{ki} - {\bf r}_{kj}, 
{\bf n}_{3i} = \frac{{\bf r}_{3i}}{r_{3i}}$ ($i$ = 1, 2) are the two unit vectors (`polaras') and $r_{ij} = 
\mid {\bf r}_{ij} \mid$ are the three relative (scalar) coordinates. In this equation and everywhere below 
the notation ${\cal Y}_{LM}^{\ell_{1} \ell_{2}}({\bf r}_{31}, {\bf r}_{32})$ stands for the bipolar harmonics 
which have been defined in \cite{Varsh} (see, also below). Note that the three relative coordinates are not 
truly independent, since the six following  `triangle' constraints $\mid r_{ik} - r_{jk} \mid \le r_{ij} \le 
r_{ik} + r_{jk}$ must always be obeyed for these coordinates. This produces a number of problems during 
careful and accurate optimization of the non-linear parameters in Eq.(\ref{exprel}). The operator 
$\hat{P}_{12}$ in Eqs.(\ref{exprel}) - (\ref{expu}) is the permutation of the two identical particles in 
symmetric three-body systems, e.g., in the $a a \mu$ ions. For such systems in Eqs.(\ref{exprel}) - 
(\ref{expu}) we have $\kappa = \pm 1$, where the final choice depends upon the parity of this bound state.  
For non-symmetric three-body systems the factor $\kappa$ equals zero identically. 

An alternative form of the exponential variational expansion is written in the three-body perimetric 
coordinates 
\begin{eqnarray}
 \Psi_{LM} = \frac{1}{2} (1 + \kappa \hat{P}_{12}) \sum_{n=1}^{N} \sum_{\ell_{1}} C_{n} 
 {\cal Y}_{LM}^{\ell_{1} \ell_{2}}({\bf r}_{31}, {\bf r}_{32}) \exp(-\alpha_{n} u_1 - \beta_{n} 
 u_2 - \gamma_{n} u_3) \; , \; \label{expu} 
\end{eqnarray}
where $u_1, u_2$ and $u_3$ are the perimetric coordinates: $u_i = \frac12 ( r_{ik} + r_{ji} - r_{jk})$ 
and $(i,j,k)$ = (1,2,3). In the both equations, Eqs.(\ref{exprel}) and (\ref{expu}) the real numbers 
$\alpha_{i}, \beta_{i}, \gamma_{i}$, where $i = 1, \ldots, N$, are the non-linear parameters of the 
exponential expansions. The choice of optimal non-linear parameters in Eqs.(\ref{exprel}) is tricky, 
but in the exponential variational expansion Eqs.(\ref{expu}) these parameters can be chosen (and then 
optimized) as arbitrary positive, real numbers and this fact substantially simplifies their accurate 
and careful optimization (see below). Note that the perimetric three-body (or triangle) coordinates 
were known to Archimedes $\approx$ 2250 years ago and Hero of Alexandria $\approx$ 2000 years ago, but 
in modern few-body physics they have been introduced by C.L. Pekeris in \cite{Pek1} (see also 
discussions in \cite{Fro2021} and \cite{MQS}). These three coordinates are simply (even linearly) 
related to the relative coordinates and vice versa:  
\begin{eqnarray}
  & & u_1 = \frac12 ( r_{21} + r_{31} - r_{32}) \; \; \; , \; \; \; r_{32} = u_2 + u_3 \; \; 
  , \; \nonumber \\
  & & u_2 = \frac12 ( r_{21} + r_{32} - r_{31}) \; \; \; , \; \; \; r_{31} = u_1 + u_3 \; \; 
  , \; \label{ucoord} \\
  & & u_3 = \frac12 ( r_{31} + r_{32} - r_{21}) \; \; \; , \; \; \; r_{21} = u_1 + u_2 \; \; 
  , \; \nonumber
\end{eqnarray}
where $r_{ij} = r_{ji}$ are the relative coordinates. Note that our definition of these (perimetric) 
coordinates is slightly different from their definition given in \cite{Pek1}.  

The three scalar, perimetric coordinates $u_1, u_2, u_3$ are very convenient for general analysis and 
highly accurate bound state numerical computations of arbitrary three-body systems. Probably, the 
perimetric coordinates $u_1, u_2$ and $u_3$ are the best choice for the three internal (or interparticle) 
coordinates in various three-body problems. This follows from the known fact that three-body perimetric 
coordinates allow one to reduce the original three-body problem to the three one-dimensional problems 
each of which can be solved (or analyzed) with the use of one perimetric coordinate only. Furthermore, 
the three perimetric coordinates $u_1, u_2, u_3$ have the following properties: (1) they are orthogonal 
to each other, (2) they are independent of each other, and (3) each of these coordinates varies between 
0 and $+\infty$. Combination of these properties of perimetric coordinates drastically simplifies 
analytical formulas for all matrix elements of the Hamiltonian and overlap matrices needed in actual 
calculations. This also simplifies numerical optimization of the non-linear parameters in the exponential 
variational expansion, Eq.(\ref{expu}) \cite{XXX}. Indeed, by applying the variational expansion 
Eq.(\ref{expu}) we do not need to check a large number of additional conditions for the chosen non-linear 
parameters, since they are always positive (in contrast with Eq.(\ref{exprel})). The same reason leads to 
a substantial simplification of analytical and numerical computations of the three-body integrals which 
are needed for solution of the corresponding eigenvalue problem and for evaluation of a large number of 
bound state properties in an arbitrary three-body system. In general, all numerical parameters in 
Eqs.(\ref{exprel}) and (\ref{expu}) can be complex, but in this study we shall not apply our exponential 
expansions written in the Fourier-like form. However, in our Appendix B we discuss a few interesting 
problems which are directly related to the Fourier form of the exponential variational expansion, 
Eq.(\ref{expu}). 

The bipolar spherical harmonics $ {\cal Y}_{LM}^{\ell_{1} \ell_{2}} ({\bf r}_{31}, {\bf r}_{32})$ 
in Eqs.(\ref{exprel}) and (\ref{expu}) are written in the form \cite{Varsh}
\begin{equation}
 {\cal Y}_{LM}^{\ell_{1},\ell_{2}}({\bf r}_{31}, {\bf r}_{32}) = r^{\ell_{1}}_{31} 
 r^{\ell_{2}}_{32} \sum_{\ell_{1}, \ell_{2}} C^{LM}_{\ell_{1} m_{1};\ell_{2} m_{2}} 
 Y_{\ell_{1} m_{1}}({\bf n}_{31}) Y_{\ell_{2} m_{2}}({\bf n}_{32}) \; , \; \label{e7}
\end{equation}
where ${\bf n}_{3k} = \frac{{\bf r}_{3k}}{r_{3k}}$ ($k$ = 1, 2) are the two unit vectors (two `polaras') 
and $C^{LM}_{\ell_{1} m_{1};\ell_{2} m_{2}}$ are the vector-coupling (or Clebsch-Gordan) coefficients 
(see, e.g., \cite{Varsh} and \cite{Rose} - \cite{FroF}). As follows from Eq.(\ref{e7}) each bipolar 
harmonic is the $M-$component of the irreducible tensor of rank $L$, where $L$ is the angular momentum 
of the given bound state. In actual bound state calculations it is possible to use only those bipolar 
harmonics for which $\ell_{1} + \ell_{2} = L$ (for all bound states of the natural parity). This 
means that each basis function in Eqs.(\ref{exprel}) and (\ref{expu}) is an eigenfunction of the 
$L^2$ and $L_z$ operators with eigenvalues $L(L+1)$ and $M$, respectively. This means that $\hat{L}^2 
\Psi_{LM} = L(L + 1) \Psi_{LM}$, while $M$ is the eigenvalue of the $\hat{L}_z$ operator, i.e.,  
$\hat{L}_z \Psi_{LM} = M \Psi_{LM}$. As is well known (see, e.g., \cite{FroSm96} and references therein) 
the angular integrals of a product of arbitrary number of bipolar harmonics can always be represented as 
a finite sum of scalar functions which depend upon the relative and/or perimetric coordinates only 
(see below). This guarantees a closure of the space of basis functions in  Eqs.(\ref{exprel}) and 
(\ref{expu}) in respect to the regular multiplication, or, in other words, it transforms this space 
into a closed Stone-Weierstrass algebra \cite{Rudin} of radial basis functions. Thus, we have shown 
that numerical computations of matrix elements of the Hamiltonian and overlap matrices is reduced to 
calculations of the corresponding radial and angular parts. Details of such calculations are considered 
in the next two Sections.      

\section{Three-body integrals in perimetric coordinates}

Let us briefly explain our approach which is extensively used to determine various radial three-body 
integrals in perimetric coordinates. As mentioned above three-body perimetric coordinates $u_1, u_2$ 
and $u_3$ have some crucial advantages in applications to a large number of three- and few-body systems. 
These advantages have been mentioned above: three perimetric coordinates $u_1, u_2$ and $u_3$ are 
independent from each other, they are orthogonal each other and each of them varies between 0 and 
$+\infty$. This explains why the perimetric coordinates are very convenient to determine various 
three-body integrals, including some very complex and singular integrals. First, consider the following 
three-body (or three-particle) integral 
\begin{eqnarray}
 & & {\cal F}_{k;l;n}(\alpha, \beta, \gamma) = 
 \int_0^{+\infty} \int_0^{+\infty} \int_{\mid r_{32} - r_{31} \mid}^{r_{32} + r_{31}} \exp[-\alpha 
 r_{32} - \beta r_{31} - \gamma r_{21}] r^{k}_{32} r^{l}_{31} r^{n}_{21} dr_{32} dr_{31} dr_{21} 
 \nonumber \\
 & & = 2 \int^{\infty}_{0} \int^{\infty}_{0} \int^{\infty}_{0} (u_3 + u_2)^{k} (u_3 + u_1)^{l} 
 (u_2 + u_1)^{n} \times \nonumber \\
 & & \exp[-(\alpha + \beta) u_3 - (\alpha + \gamma) u_2 - (\beta + \gamma) u_1] du_1 du_2 du_3 \; 
 \; , \; \label{mine}
\end{eqnarray}
where all indexes $k, l, n$ are integer and non-negative, while 2 is the Jacobian ($\frac{\partial 
r_{ij}}{\partial u_{k}}$) of the $r_{ij} \rightarrow u_k$ substitution. This integral is called 
the fundamental three-body integral, since the knowledge of the ${\cal F}_{k;l;n}(\alpha, \beta, 
\gamma)$ function allows one to determine a large number of different three-body integrals, which 
are needed for solution of the original Schr\"{o}dinger equation $\hat{H} \Psi = E \Psi$ for a 
given three-body system. Applications and high efficiency of the three perimetric coordinates $u_1, 
u_2$ and $u_3$ can be demonstrated by derivation of the closed analytical formula for the integral, 
Eq.(\ref{e11}). Here we just present the final result. The explicit formula for the ${\cal 
F}_{k;l;n}(\alpha, \beta, \gamma)$ integral is written in the form
\begin{eqnarray}
 {\cal F}_{k;l;n}(\alpha, \beta, \gamma) &=& 2 \sum^{k}_{k_1=0} \sum^{l}_{l_1=0} \sum^{n}_{n_1=0} 
 C_{k}^{k_1} C_{l}^{l_1} C_{n}^{n_1} \frac{(l-l_1+k_1)!}{(\alpha + \beta)^{l-l_1+k_1+1}} 
 \frac{(k-k_1+n_1)!}{(\alpha + \gamma)^{k-k_1+n_1+1}} \times \nonumber \\
 & & \frac{(n-n_1+l_1)!}{(\beta + \gamma)^{n-n_1+l_1+1}} \; \; \; , \; \label{e11} 
\end{eqnarray}
where $C^{m}_{M}$ is the number of combinations from $M$ by $m$ (here $m$ and $M$ are integer non-negative 
numbers). The formula, Eq.(\ref{e11}), can also be written in a few different (but equivalent!) forms. For 
the first time this formula, Eq.(\ref{e11}), was derived by me in the middle of 1980's \cite{Fro87}. As 
mentioned above the ${\cal F}_{k;l;n}(\alpha, \beta, \gamma)$ integrals play a central role in physics of 
three-body systems. 

The second fundamental integral is a direct generalization of the integral, Eq.(\ref{e11}), and it contains 
the both relative and perimetric coordinates. This `mixed' integral is written in the form   
\begin{eqnarray}
 & &{\cal H}^{p;q;t}_{k;l;n}(\alpha, \beta, \gamma; \lambda, \mu, \nu) = 
 2 \int_0^{+\infty} \int_0^{+\infty} \int_{\mid r_{32} - r_{31} \mid}^{r_{32} + r_{31}} \exp[-\alpha r_{32} 
 - \beta r_{31} - \gamma r_{21} - \nu u_3 - \mu u_2 - \lambda u_1] \times \nonumber \\ 
 & &r^{k}_{32} r^{l}_{31} r^{n}_{21} u^{p}_{1} u^{q}_{2} u^{t}_{3} dr_{32} dr_{31} dr_{21}= 2 
 \int^{\infty}_{0} \int^{\infty}_{0} \int^{\infty}_{0} (u_3 + u_2)^{k} u^{p}_{1} (u_3 + u_1)^{l} u^{q}_{2} 
 (u_2 + u_1)^{n} u^{t}_{3} \times \nonumber \\ 
 & & \exp[-(\alpha + \beta + \nu) u_3 - (\alpha + \gamma + \mu) u_2 - (\beta + \gamma + \lambda) u_1] du_1 
 du_2 du_3 \; \; \; . \; \label{mine} 
 \end{eqnarray} 
Analytical formula for this integral takes the form 
\begin{eqnarray}
 & & {\cal H}^{p;q;t}_{k;l;n}(\alpha, \beta, \gamma; \lambda, \mu, \nu) = 2 \sum^{k}_{k_1=0} \sum^{l}_{l_1=0} 
 \sum^{n}_{n_1=0} 
 C^{k_1}_{k} C^{l_1}_{l} C^{n_1}_{n} \label{e110} \\
 & & \frac{(k + l - l_1 + t)! (k - k_1 + n_1 + q)!  (l_1 + n - n_1 + p)!}{(\alpha + \beta + \nu)^{k+l-l_1+t+1} 
 (\alpha + \gamma +  \mu)^{k-k_1+n_1+q+1} (\beta + \gamma + \lambda)^{l_1+n-n_1+p}} \nonumber 
\end{eqnarray}
and it is also clear that ${\cal H}^{0;0;0}_{k;l;n}(\alpha, \beta, \gamma; 0, 0, 0) = {\cal F}_{k;l;n}(\alpha, 
\beta, \gamma)$. Again, we have to note that the formula, Eq.(\ref{e110}), can also be written in a number 
of different (but equivalent) forms. Furthermore, any direct derivation of this formula in the relative 
three-body coordinates is quite difficult. Note also that in this study we are not dealing with any singular, 
quasi-singular and other `special' three-body integrals which will be discussed elsewhere.   

The analytical formulas, Eqs.(\ref{e11}) and (\ref{e110}), for the two fundamental integrals are relatively 
simple and they do not lead to any numerical instabilities which can restrict actual computations of matrix 
elements of the Hamiltonian and overlap matrices. These formulas allow one to develop a number of fast, 
numerically stable and relatively simple algorithms which work very well in bound state computations of 
various three-body systems. Highly accurate results obtained with the use of formulas derived in this Section 
for the weakly-bound (1,1)-states and rotationally excited (2,0)-states in the three-body muonic molecular 
$d d \mu, d t \mu$ and $t t \mu$ ions are discussed below.    

\section{Angular parts of three-body integrals}

In the previous Section we have derived and discussed the two analytical formulas for the radial integrals 
which are often used in actual bound state calculations. Now, we want to discuss analytical formulas which 
have been derived to determine the corresponding angular integrals. This is the second part of our 
computational method which is equally important to develop the closed variational procedure for highly 
accurate, bound state computations of many three-body systems. Note that each basis function in 
Eqs.(\ref{exprel}) and (\ref{expu}) includes one bipolar harmonic. Therefore, in order to perform numerical 
computations of the rotationally excited bound states in muonic molecular ions we need to derive the closed 
analytical formulas for the products of different numbers of bipolar harmonics. In fact, for three-body 
systems with the scalar interactions between particles, e.g., for all Coulomb three-body systems, one always 
needs formulas which include the products of two bipolar harmonics only. In such cases the corresponding 
angular integrals are written in the form: 
\begin{eqnarray}
 \oint \Bigl[ {\cal Y}_{LM}^{\ell_{1} \ell_{2}}({\bf n}_{31}, {\bf n}_{32}) \Bigr]^{\ast} {\cal 
  Y}_{LM}^{\ell_{3} \ell_{4}}({\bf n}_{31}, {\bf n}_{32}) d{\bf n}_{31} d{\bf n}_{32} = 
  f^{L}_{\ell_1\ell_2;\ell_3\ell_4}(r_{32}, r_{31}, r_{21}) = \sum_{K} C_{K} P_{K}(\cos\theta) \; , 
  \; \label{angint1} 
\end{eqnarray}
where $\cos \theta = \cos\theta_{31^{\wedge}32} = \frac{{\bf r_{31}} \cdot {\bf r_{32}}}{r_{31} \; r_{32}} 
= \frac{r^{2}_{31} + r^{2}_{32} - r^{2}_{21}}{2 \; r_{31} \; r_{32}}$, while the coefficients $C_K$ are
\begin{eqnarray}
 C_K &=& \frac12 (-1)^{L+K} \sqrt{[\ell_1] [\ell_2] [\ell_3] [\ell_4]} \; [K] \;
    \left( \begin{array}{ccc} \ell_1 & \ell_3 & K \\ 0 & 0 & 0 \end{array} \right) 
    \left( \begin{array}{ccc} \ell_2 & \ell_4 & K \\ 0 & 0 & 0 \end{array} \right) 
   \left\{ \begin{array}{ccc} \ell_3 & \ell_4 & L \\ \ell_2 & \ell_1 & K \end{array} \right\}
   \; , \; \label{angint2}
\end{eqnarray}
where $[ a ] = 2 a + 1$ and the notation $P_{K}(x)$ in Eq.(\ref{angint1}) stands for the Legendre polynomial of 
order $K$, where $K$ is a non-negative integer. These polynomials can be determined by using the well known 
recursion formula: $(n + 1) P_{n+1}(x) = (2 n + 1) x P_{n}(x) - n P_{n-1}(x)$ and the two following `boundary 
conditions': $P_{0}(x) = 1$ and $P_{1}(x) = x$ \cite{GR}. Also, in the formula, Eq.(\ref{angint2}), we use the 
standard notations for the $3 j$- and $6 j$-symbols \cite{Edm}. The sum over $K$ in Eq.(\ref{angint1}) is always 
finite, since the product of two $3 j-$symbols is not zero only for those integer values $K$ which are bounded 
between the two following values: $max\{| \ell_1 - \ell_3 |, | \ell_2 - \ell_4 |\}$ and $min \{ \ell_1 + \ell_3, 
\ell_2 + \ell_4 \}$, i.e., $max\{| \ell_1 - \ell_3 |, | \ell_2 - \ell_4 |\} \le K \le min \{ \ell_1 + \ell_3, 
\ell_2 + \ell_4 \}$. Moreover, the product of these two $3 j-$symbols equals zero unless the two following sums 
of the corresponding momenta: $\ell_1 + \ell_3 + K$ and $\ell_2 + \ell_4 + K$ are even numbers. Briefly, this 
means that the $f^{L}_{\ell_1\ell_2;\ell_3\ell_4}(r_{32}, r_{31}, r_{21})$ factor in Eq.(\ref{angint1}) is a 
scalar, polynomial function of the three relative coordinates $r_{32}, r_{31}$ and $r_{21}$ only, i.e., it does 
not include any free angular factor. 

In some three-body papers, including our earlier papers, the angular integrals, which include products of two 
bipolar harmonics $\Bigl[ {\cal Y}_{LM}^{\ell_{1},\ell_{2}}({\bf r}_{31}, {\bf r}_{32}) \Bigr]^{\ast} {\cal 
Y}_{LM}^{\ell_{3},\ell_{4}}({\bf r}_{31}, {\bf r}_{32}) \simeq (-1)^L {\cal Y}_{LM}^{\ell_{1},\ell_{2}}({\bf 
r}_{31}, {\bf r}_{32}) {\cal Y}_{LM}^{\ell_{3},\ell_{4}}({\bf r}_{31}, {\bf r}_{32})$ have also been considered. 
In these cases our formula, Eq.(\ref{angint1}), is written in the same form, but with a different radial factor 
$F^{L}_{\ell_1\ell_2;\ell_3\ell_4}(r_{32}, r_{31}, r_{21}) = r^{\ell_1+\ell_3}_{31} r^{\ell_2+\ell_4}_{32} 
f^{L}_{\ell_1\ell_2;\ell_3\ell_4}(r_{32}, r_{31}, r_{21})$. In these cases the formula, Eq.(\ref{angint1}), 
takes the form  
\begin{eqnarray}
 & & \oint \Bigl[ {\cal Y}_{LM}^{\ell_{1} \ell_{2}}({\bf r}_{31}, {\bf r}_{32}) \Bigr]^{\ast} {\cal 
 Y}_{LM}^{\ell_{3} \ell_{4}}({\bf r}_{31}, {\bf r}_{32}) d{\bf n}_{31} d{\bf n}_{32} = 
 F^{L}_{\ell_1\ell_2;\ell_3\ell_4}(r_{32}, r_{31}, r_{21}) \nonumber \\
 &=& r^{\ell_1+\ell_3}_{31} r^{\ell_2+\ell_4}_{32} f^{L}_{\ell_1\ell_2;\ell_3\ell_4}(r_{32}, 
 r_{31}, r_{21}) = r^{\ell_1+\ell_3}_{31} r^{\ell_2+\ell_4}_{32} \sum_{K} C_{K} 
 P_{K}\Bigl(\frac{r^{2}_{31} + r^{2}_{32} - r^{2}_{21}}{2 r_{31} r_{32}}\Bigr) \; \; . \; \label{angint10} 
\end{eqnarray}
The fact that the both `radial' $F^{L}_{\ell_1\ell_2;\ell_3\ell_4}(r_{32}, r_{31}, r_{21})$ and 
$f^{L}_{\ell_1\ell_2;\ell_3\ell_4}(r_{32}, r_{31}, r_{21})$ factors in the equations, Eqs.(\ref{angint1}) 
and (\ref{angint10}), are the scalar functions of the three relative coordinates $r_{32}, r_{31}$ and 
$r_{21}$ only is crucial to develop the closed analytical procedure for determining all matrix elements 
which are needed in calculations of the bound states in three-body Coulomb systems with arbitrary, in 
principle, angular momentum $L$ and its $z-$projection $M$. Furthermore, as follows from the formulas, 
Eq.(\ref{angint1}) and (\ref{angint10}), the basis functions from the both exponential expansions, 
Eqs.(\ref{exprel}) and (\ref{expu}), form the closed Stone-Weierstrass algebra of functions \cite{Rudin}, 
which are defined either in the three-dimensional space of relative coordinates $R_{ijk}$ space, or in 
three-dimensional space of perimetric coordinates $U_{ijk}$, which is, in fact, the direct product of 
three one-dimensional $U_i-$spaces, i.e., $U_{ijk} = U_{1} \otimes U_{2} \otimes U_{3}$ (see above).    

\section{Numerical calculations and results}

As mentioned above all calculations of the bound (1,1)- and (2,0)-states of muonic molecular ions in this study 
have been performed in muon-atomic unis. The masses of particles used in our computations are:
\begin{eqnarray}
 &&M_{t} = 5496.92153573 \; \; m_e \; , \; m_{\mu} = 206.7682830 \; \; m_e \; , \; \nonumber \\
 &&M_{d} = 3670.48296788  \; \; m_e \; , \;  M_{p} = 1836.15267343 \; \; m_e \; . \; \label{masses} 
\end{eqnarray}  
These masses of the negatively charged muon $\mu^{-}$ and nuclei of hydrogen isotopes are currently recommended 
for scientific use by CODATA/NIST in 2022 \cite{NIST}. In general, it is very convenient to perform numerical 
calculations of the bound states in muonic ions by using the muon-atomic units, where the masses of particles, 
Eq.(\ref{masses}), are $m_{\mu} = 1, M_{i} = \frac{M_{i}}{m_{\mu}}$ and $i = d, t$. The main advantage of the 
muon-atomic units ($\hbar = 1, m_{\mu} = 1$ and $e = 1$) is obvious, since in these units we are not losing 
accuracy in operations with additional physical factors and constants each of which is usually known to a 
restricted numerical accuracy. Therefore, it always better to perform all `direct' variational calculations in 
muon-atomic units. However, when these calculations were finished, then it is easy to recalculate results to 
the atomic units, electron-Volts, and other similar energy units. For instance, for muonic molecular three-body 
$(a a \mu)^{+}$ and $(a b \mu)^{+}$ ions there is an old tradition to express their total and binding energies 
in electron-Volts. To recalculate our variational energies from muon-atomic units to $eV$ one needs to apply the 
doubled Rydberg constant which equals $2 Ry$ = 27.211386028 $eV$ and 1 $m.a.u.$ = 206.7682830 $a.u.$ The total 
factor used in similar re-calculations is $2 Ry \cdot m_{\mu}$ = 5626.451612133 $eV (m.a.u.)^{-1}$. Note that 
the doubled Rydberg constant is closely related to the rest mass of the electron $m_e$ = 0.510 998 9461 
$MeV/c^{2}$. Indeed, the well known condition $2 Ry = \alpha^{2} m_e c^{2}$ must always be obeyed. Here 
$\alpha$ is the dimensionless fine-structure constant $\alpha$ = 7.2973525693 $\cdot 10^{-3}$ and $c = 
2.99792458 \cdot 10^{10}$ $cm \cdot sec^{-1}$ is the speed of light in vacuum. 

Results of our highly accurate, variational calculations of the total energies $E$ for the weakly-bound 
(1,1)-states in the muonic $d d \mu $ and $d t \mu$ ions and rotationally excited (2,0)-states in the $d d \mu, 
d t \mu$ and $t t \mu$ ions can be found in Table I. All results in Table I are presented in muon-atomic units 
and for different number(s) $N$ of basis functions used in Eq.(\ref{expu}). As follows from Table I our current 
numerical accuracy achieved in variational computations of the weakly-bound (1,1) states in the $d t \mu$ and 
$d d \mu$ muonic ions is very high. Note that our new results are significantly more accurate (approximately 
in $1 \cdot 10^{3}$ times) than similar results obtained for these weakly-bound states a few years ago 
\cite{FroPLA}. This means that currently we can provide three exact decimal digits extra which have been 
stabilized in our calculations of the total energies. 

In old papers (see, e.g., \cite{Vinn}, \cite{BD1}, \cite{Kam}) the final results usually included only the 
binding $\varepsilon$ energies, which were usually expressed in the electron-Volts. In general, for some 
bound state ($BS$) in arbitrary three-body muonic ion $a b \mu$, where $m_a \le m_b$, the binding energy 
$\varepsilon_{BS}$ is determined as the difference between the total energy $E_{BS}(a b \mu)$ of this state 
and the total energy of the ground $S(L = 0)-$state in the two-body, muonic $b \mu$ atom. In electron-Volts 
the binding energy of the muonic $a b \mu$ ion is written in the form    
\begin{eqnarray}
 \varepsilon_{BS}(a b \mu) &=& 2 Ry \; m_{\mu} \; \Bigl[ E_{BS}(a b \mu) - \frac{1}{2 \Bigl(1 + 
 \frac{m_{\mu}}{m_{b}} \Bigr)} \Bigr] = 5626.451612133 \times \; \nonumber \\ 
 & & \Bigl[ E_{BS}(a b \mu) - \frac{1}{2 \Bigl(1 + \frac{m_{\mu}}{m_{b}}\Bigr)} \Bigr] \; \; eV \; , \; 
 \label{bin-enrgy} 
\end{eqnarray} 
where $E_{BS}(a b \mu)$ are the total energies of muonic molecular ions presented in Table I. The second term 
in each of the equations, Eqs.(\ref{bin-enrgy}), is the total energy of the ground $S(L = 0)-$state of the 
neutral muonic atom $b \mu = b^{+} \mu^{-}$ expressed in muon-atomic units. The formula, Eq.(\ref{bin-enrgy}), 
allows one to obtain the corresponding binding energies of all muonic molecular ions considered in this study 
in electron-Volts. For instance, our best variational binding energies of the weakly-bound (1,1)-states in 
the muonic $d d \mu$ and $d t \mu$ ions can be evaluated as $\approx$ -1.97498052811372759 $eV$ and $\approx$ 
-0.66032991050518274 $eV$, respectively. Analogous binding energies of the rotationally excited (bound) 
(2,0)-states in the muonic $d d \mu, d t \mu$ and $t t \mu$ ions are: $\approx$ -102.648490859750 $eV$, 
-86.4932595208150 $eV$ and -172.701007428930 $eV$, respectively. In recent scientific papers the binding energies 
of muon molecular ions are expressed in $eV$ only occasionally. 
 
In general, the dimensionless ratio $\tau = \frac{\varepsilon_{BS}(a b \mu)}{E_{BS}(a b \mu)}$ can be used as 
a `criterion of boundness' of the given $BS-$state in the muonic $a b \mu$ ion. For very weakly-bound states in 
the muonic $d t \mu$ and $d d \mu$ ions the numerical values of this parameter $\tau$ are extremely small: $\tau 
\approx$ 0.0002435 (or 0.02435\%) and $\tau \approx$ 0.0007410 (or 0.07410\%), respectively. Probably, these two 
ratios $\tau$ are the smallest among all stable three-particle systems known in nature (here we do not want to 
discuss the highly excited Rydberg states in the helium-like atoms and ions, see, e.g., \cite{FroF} and 
references therein). In fact, it was noticed already in \cite{Fro1992} (see also earlier references therein) that 
for all Coulomb three-body systems with unit charges the ratios $\tau$ do not exceed 0.17 (or 17\%) \cite{YYY}. 
For the ground (bound) states in the muonic $a b \mu$ and $a a \mu$ ions, where $(a,b) = (p, d, t)$, the ratio 
$\tau$ does not exceed 0.07 - 0.08, while for all rotationally and vibrationally excited states these ratios are 
much smaller, e.g., $\tau \le 0.01 - 0.02$. In other words, all bound states in the three-body muonic $a b \mu$ 
and $a a \mu$ ions are relatively weakly-bound, since for all these states the ratios $\tau$ of their binding and 
total energies $\tau = \frac{\varepsilon}{E}$ are small.

In this study we also determine a number of bound state properties of the muonic molecular $d d \mu, d t \mu$ 
and $t t \mu$ ions in their (1,1)- and (2,0)-states which are of great interest in applications. The properties 
of these bound states have never been determined in earlier studies even to a relatively good accuracy. There 
were only a very few attempts to evaluate such properties numerically by using some model, very approximate wave 
functions. Briefly, it is possible to say that these properties are currently unknown. In general, any highly 
accurate computations of properties of the bound states in muonic molecular ions with $L \ge 1$ are significantly 
more complicated than analogous computations for the $S(L = 0)-$states \cite{Fro2011}. For the weakly-bound 
(1,1)-states (or $P^{*}(L = 1)$ states) there are some additional complications. 

Now, let us define the bound state property in some quantum system. In general, for an arbitrary bound state in 
some quantum few-body system we can determine the following expectation value 
\begin{eqnarray}
  X = \frac{\langle \Psi \mid \hat{X} \mid \Psi \rangle}{\langle \Psi \mid \Psi \rangle} = 
  \langle \Psi_N \mid \hat{X} \mid \Psi_N \rangle \; , \; \label{prop1}
\end{eqnarray}
where $\langle \Psi_N \mid \Psi_N \rangle = 1$ and $\Psi_N$ is the unit-norm wave function. This expectation value 
is called the bound state property $X$ of this system in a given bound state \cite{Dirac}. This property is 
uniformly determined by the given self-adjoint operator $\hat{X}$ and the bound state wave function $\Psi$. As is 
well known (see, e.g., \cite{Dirac}) the bound state wave functions always have the finite norms. Therefore, below 
we shall assume that the wave function $\Psi$ has the unit norm, i.e., in Eq.(\ref{prop1}) $\Psi = \Psi_N$. In this 
study we deal with the muonic molecular ions which are the Coulomb three-body system. Therefore, we can assume that 
the operator $\hat{X}$ in Eq.(\ref{prop1}) is an explicit function of the three relative $r_{ij}$, or three 
perimetric coordinates $u_{k}$, where $(i,j,k) = (1,2,3),$ and momenta of the three particles ${\bf p}_i$. By 
choosing different operators $\hat{X}$ in Eq.(\ref{prop1}) we obtain different expectations values, or bound states 
properties. 

A number of bound state properties of the (1,1)- and (2,0)-states in the muonic $d t \mu$ ion can be found in 
Tables II and III, respectively. All these properties are expressed in muon-atomic units. Physical meaning of 
each of these bound state properties follows from the notation used in Tables II and III. For instance, the 
$\langle r_{ij} \rangle$ means the expectation values of the linear distance between particles $i$ and $j$. 
Analogously, the expectation values $\langle r^{k}_{ij} \rangle$ mean the $k-$th powers of these interparticle 
distances. In this paper we consider the cases when $k$ = -1, 1, 2 and 3. In the non-symmetric muonic $a b 
\mu$ ion it is convenient to designate the particles by numbers 1, 2 and 3, where the particle 3 is the 
negatively muon $\mu^{-}$, while the particles 1 and 2 are the nuclei of two hydrogen isotopes, i.e., $d$ and 
$t$, respectively. The results from Table II indicate clearly that the (1,1)-state in this ion is a truly 
weakly-bound state which has the cluster two-body structure which corresponds to actual pre-dissociation of 
this ion. Briefly, this means that in the weakly-bound (1,1)-state the deuterium nucleus moves as a single 
particle in the `effective' field of the `central' muonic atom $t \mu$. In other words, the structure of the 
weakly-bound (1,1)-state can approximately be represented by the equation $(d t \mu)^{+} \approx d^{+} + t 
\mu$. Indeed, as follows from Table II the interparticle tritium-muonic distance equals $\langle r_{t \mu} 
\rangle \approx 1.20$ $m.a.u.$ This numerical value is close to the tritium-muonic distance in the two-body 
muonic $t \mu$ atom. Furthermore, the both $\langle r_{d \mu} \rangle$ and $\langle r_{d t} \rangle$ 
distances, are significantly larger than the corresponding tritium-muon distance $\langle r_{t \mu} \rangle$. 
Analogous relations are obeyed for higher powers of these inter-particle distances, i.e., $\langle r^{k}_{d 
\mu} \rangle \gg \langle r^{k}_{t \mu} \rangle$ and $\langle r^{k}_{d t} \rangle \gg \langle r^{k}_{t \mu} 
\rangle$, where $k$ is integer and $k \ge 2$. 

The bound state properties of the (2,0)-state in the non-symmetric $d t \mu$ ion are shown in Table III. 
As follows from Table III these properties correspond to some `regular' state, which is not extremely 
weakly-bound as the (1,1)-state in the same ion discussed above (see Table II). In particular, in this ion 
all positive powers of the deuterium-muon and tritium-muon distances are noticeably smaller than the 
corresponding powers of deuterium-tritium distance(s) $\langle r^{k}_{d t} \rangle$. This essentially means 
that the bound (2,0)-state in the muonic $d t \mu$ ion is relatively weakly-bound, but the negatively 
charged muon $\mu^{-}$ is not localized almost completely on the tritium nucleus (compare with the 
(1,1)-state in this ion discussed above). A very similar situation can be found in the bound (2,0)-states 
of the both muonic $d d \mu$ and $t t \mu$ ions. The bound state properties of these ions are presented in 
Table IV. These two states are also relatively weakly-bound, but not as as extreme as the (1,1)-state in 
the muonic $d d \mu$ ion (see Table V below). 

Now, let us discuss the bound state properties of the weakly-bound (1,1)-state in the symmetric muonic $d 
d \mu$ ion, which can be found in Table V. It is clear $a$ $priory$ that such bound state properties of the 
both weakly-bound (1,1)-states in the non-symmetric $d t \mu$ and symmetric muonic $d d \mu$ ions must be 
quite close (numerically) to each other. This follows from the fundamental physical principles and continuity 
of the solution(s) of the Schr\"{o}dinger equation upon a few physical parameters, e.g., the masses of 
particles. In turn, the last statement is based on the well known Poincare theorem (see, e.g., \cite{Migd}). 
However, in respect to the current methodology widely accepted in modern Quantum Mechanics we cannot report 
separately the two different $\langle r^{k}_{31} \rangle$ and $\langle r^{k}_{32} \rangle$ values, since they 
both mean the same $\langle r^{k}_{d \mu} \rangle$ expectation value. In other words, the expectation value 
$\langle r^{k}_{d \mu} \rangle = \frac12 \Bigl(\langle r^{k}_{31} \rangle + \langle r^{k}_{32} \rangle\Bigr)$
must always be symmetrized upon the two identical (or indistinguishable) deuterium nuclei, or particles 1 and 
2 in our current notation. On the other hand, this expectation value is essentially useless for investigation 
of the actual structure of any weakly-bound state with the two-body cluster structure which corresponds to   
actual pre-dissociation. 

Suppose we have a short wavelength microscope to observe the internal structure of some extremely weakly-bound 
state in the symmetric muonic $d d \mu$ (= $d^{+} d^{+} \mu^{-}$) ion. In this microscope, we will see that 
one deuterium nucleus $d^{+}$ always moves closely to the negatively charged muon $\mu^{-}$, while the second 
deuterium nucleus $d^{+}$ always moves at a significant distance from the two-particle, neutral cluster 
$d^{+}\mu^{-}$. Here is important to note that we are dealing with the true weakly-bound state and the distance 
between two deuterium nuclei is really large. Therefore, in this case we can define the outermost (from muon) 
and closest (to muon) deuterium nuclei. Below (see, also Table V), the outermost deuterium nucleus is designated 
by the index $o$, while the closest deuterium nucleus is denoted by the index $c$. Note that the two nuclei in 
the $d d \mu$ ion are identical and Pauli indistinguishability principle is applied to them. This means that we 
cannot say where is the nucleus 1, or nucleus 2, since they are indistinguishable nuclei (or `particles'). 
However, from actual optical experiments one can always determine the locations of the outermost and closest 
deuterium nuclei. Briefly, if we are using the indexes $c$ and $o$ to label our expectation values, then we can 
avoid any forced symmetrization of these expectation values. The advantage of the new system of notation can 
exactly be seen from Table V, where one finds, $\langle r_{c \mu} \rangle \approx$ 1.0625016 $m.a.u.$, which is 
numerically close to the corresponding expectation $\langle r_{32} \rangle$ value (1.1997897 $m.a.u.$) obtained 
for the weakly-bound (1,1)-state of the $d t \mu$ ion (see Table II). Similar `numerical closeness' can also be 
found between the $\langle r^{k}_{c \mu} \rangle$ and $\langle r^{k}_{32} \rangle$ expectation values, where $k 
\ge 2$, from Table V (symmetric $d d \mu$ ion) and Table II (non-symmetric $d t \mu$ ion), respectively. 

\section{Conclusions}

We have performed highly accurate numerical computations of the weakly-bound (1,1)-states in the three-body 
muonic $d d \mu $ and $d t \mu$ ions and rotationally excited (2,0)-states in the muonic $d t \mu, d d \mu$ 
and $t t \mu$ ions. Based on recently developed and effective optimization strategy of the non-linear 
parameters in our trial variational wave functions we could determine the both total and binding energies of 
the weakly-bound (1,1)-states in the $d d \mu $ and $d t \mu$ muonic ions to very high numerical accuracy. 
Such an accuracy is substantially better than analogous accuracy known from earlier computations of the 
weakly-bound (1,1)-states in these muonic ions. In fact, our current accuracy for these weakly-bound, excited 
states is quite comparable with the maximal numerical accuracy achieved in calculations of other bound states 
in the muonic molecular ions, including the corresponding ground bound states.   
  
We have also developed some accurate numerical method which can be used to determine different bound state 
properties of the weakly-bound (1,1)-states in the three-body $d d \mu $ and $d t \mu$ muonic ions and 
rotationally excited (2,0)-states in the $d d \mu, d t \mu$ and $t t \mu$ ions. Results of our accurate 
computations of these properties are presented and discussed in this study. None of the bound state properties 
have ever been evaluated (to high accuracy) for these bound states in the muonic $d d \mu , d t \mu$ and $t t 
\mu$ ions. To describe the pre-dissociation two-body structure of the weakly-bound $(1,1)-$state in the 
symmetric $d d \mu $ ion we introduced the new system of notations which is based on an explicit designation 
of the outermost $o$ and closest $c$ deuterium nuclei in the $d d \mu$ ion. It appears that this new system of 
notation can also be very effective to describe the structure of weakly-bound and cluster bound states in all 
symmetric three-body systems, including some negatively charged ions, the excited states of different neutral 
and positively charged atomic atoms and ions, etc. In the following Appendix A we discuss the current status 
of muon-catalyzed nuclear fusion. In our second Appendix B we develop the new universal variational expansion 
for three-body systems. This new expansion can be used for extremely accurate bound state calculations of 
arbitrary three-body systems (no mass-limits), including the truly adiabatic ${}^{\infty}$H$^{+}_{2}$ ion and 
close systems. 

\appendix
\section{On resonance muon-catalyzed nuclear fusion}
\label{A} 

Let us briefly discuss the role which the both weakly-bound (1,1)-states in the muonic molecular $d d \mu$ and 
$d t \mu$ ions play in the resonance muon-catalyzed nuclear fusion in liquid hydrogen, deuterium and different 
deuterium-tritium mixtures, including equimolar (or 1:1) DT mixture. The general idea of muon-catalyzed fusion 
in liquid hydrogen was proposed in late 1940's (see, e.g., \cite{Frank}). It was assumed that one negatively 
charged, fast muon $\mu^{-}$ can repeatedly form muonic molecular ions in liquid hydrogen/deuterium mixtures 
and produce reactions of nuclear fusion in these ions. In the middle of 1950's it was well understood that this 
process includes a number of important steps. At the first stage, one muon $\mu^{-}$ will slow down in liquid 
hydrogen/deuterium and form the neutral muonic-atom, e.g., $d \mu$, which is a very compact and neutral two-body 
system. The effective diameter (spatial size) of the $p \mu, d \mu, t \mu$ quasi-atoms is $\approx$ 206 times 
smaller than typical sizes of any regular atomic system. This means that each of the two-body muonic $a \mu$ ions 
is a very compact and neutral system. Polarizabilities of these muonic quasi-atoms by usual inter-molecular and 
inter-atomic forces are also very small. Formally, each of these muonic two-body $p \mu, d \mu, t \mu$ quasi-atoms 
can be considered as an atomic `neutron', which (due to its neutrality and compactness) can freely move through 
various atoms, molecules and  molecular ions. 

At the second stage of the process, this muonic atom forms some three-body molecular ion, e.g., $p d \mu$, 
which is a part of the six-particle, two-center (or molecular) complex $[(p d \mu)^{+} p^{+}] e^{-}_2$. The 
positively charged $p d \mu$ ion plays a role of one of the two heavy nucleus in this neutral $[(p d \mu)^{+} 
p^{+}] e^{-}_2$ molecule. Eventually, in this very compact $(R \approx 2.5 \cdot 10^{-11}$ $cm$), three-particle 
$p d \mu$ ion the reaction of nuclear fusion $p + d = {}^{3}$He$ + \gamma$ occurs. After such a fusion reaction 
the negatively charged $\mu^{-}$ muon becomes free and relatively fast. It slows down in liquid hydrogen 
(deuterium, etc), forms the new three-body molecular ion, e.g., $d d \mu$ and the chain of processes/reactions 
repeats again and again. Roughly speaking, one negatively charged muon `catalyzes' a number of fusion reactions. 
Formally, the total number of fusion is restricted only by a relatively short life-time $\tau \approx 2.03 \cdot 
10^{-6}$ $sec$ of the negatively charged muon $\mu^{-}$.   

In the middle of 1950's there were a large number of speculations about how many fusion reactions can be 
catalyzed by a single muon. Finally, in 1957 Alvaretz et al \cite{Alv} performed an experiment in liquid hydrogen 
(protium) which also contained a small admixture of deuterium. It was shown in these experiments that the actual 
number of fusion reactions (per muon) $\kappa$ in the muonic molecular $p d \mu$ ions is small and usually varies 
between 1 and 2. Other muonic molecular ions were not considered in \cite{Alv}. After publication of experimental 
results in \cite{Alv} scientific community lost any interest to the muon-catalyzed nuclear fusion in liquid hydrogen 
for quite some time. 

It is interesting to investigate the reasons why traditional muon-catalyzed nuclear fusion fails in the molecular
hydrogen. In a number of papers it was found that the slowest stage of muon-catalyzed fusion is the formation of 
three-body muon molecular ions, e.g., $p d \mu, d d \mu, d t \mu$ and other similar ions. The main problem here 
is a very substantial amount of energy ($\approx$ 100 - 300 $eV$) which is released during actual formation of 
such ions. To obey the energy conservation law in the six-particle molecular complex $[(p d \mu)^{+} p^{+}] 
e^{-}_2$ this additional energy must be distributed somehow between other particles included in this molecular 
complex, i.e., between the protium nucleus and two electrons. The leading process here is the molecular 
ionization, e.g., for the $[(p d \mu)^{+} p^{+}] e^{-}_2$ molecule  
\begin{eqnarray}
 d \mu + {\rm H}_2 = [p (p d \mu)] e_2 = \Bigl\{[p (p d \mu)] e \Bigr\}^{+} + e^{-} \; \; , \; 
 \label{photo}
\end{eqnarray} 
where the kinetic energy of the final (free and fast) electron is determined from the energy conservation law. 
However, this reaction proceeds very slow, since electron ionization of the six-body molecule takes a long time. 
Indeed, the overall probability of the process, Eq.(\ref{photo}), is proportional to the fine-structure constant 
$\alpha \approx \frac{1}{137}$ which is a small parameter. After such an ionization the nuclear fusion proceeds 
in the five-body, two-center (or molecular) $\Bigl\{[p (p d \mu)] e \Bigr\}^{+}$ ion. 

Another possible process which can be used to form the three-body muonic molecular ion, e.g., $p d \mu$, is the 
regular dissociation of the $[(p d \mu)^{+} p^{+}] e^{-}_2$ molecule
\begin{eqnarray}
 d \mu + {\rm H}_2 = [p (p d \mu)] e_2 = [(p d \mu)] e  + (p e) \; \; \; . \;  \label{diss}
\end{eqnarray} 
In this case the energy released during formation of the bound muonic molecular ion is distributed between the 
two final $[(p d \mu)] e$ and $(p e)$ hydrogen-like atoms. The nuclear fusion proceeds in the neutral four-body 
$[(p d \mu)] e$ atom. This reaction also proceeds very slow, since its overall probability is proportional to 
the product of the fine-structure constant $\alpha \approx \frac{1}{137}$ and electron-nuclear mass ratio 
$\frac{m_e}{M} \le \frac{1}{1836}$. All other similar processes also have small (or very small) probabilities.  

Nevertheless, a few years after publication \cite{Alv} a group of young (at that time) fellows from Dubna (Russia) 
\cite{Bel} has decided to investigate the bound state spectra in all six muonic molecular ions: $p p \mu, p d \mu, 
p t \mu, d d \mu, d t \mu$ and $t t \mu$. They have found 20 bound states, but the two (1,1)-states in the both 
symmetric $d d \mu$ and non-symmetric $d t \mu$ ions could not be identified as the truly bound states, since their 
binding energies (if they are bound) are very small. In fact, the binding energies of these two (1,1)-states are 
less than the `threshold energy' 4.5 $eV$ which approximately equals to the dissociation energies of the neutral 
hydrogen molecules: H$_{2}$, D$_{2}$, DT, T$_{2}$, etc. Stability of these weakly-bound states could open a new 
road to the so-called `resonance' muon-catalyzed nuclear fusion. In such a resonance muon-catalyzed nuclear fusion 
the energy released during formation of weakly bound (1,1)-state in the $d t \mu$ ion (for the $d d \mu$ ion 
analogously) is instantly transferred into molecular excitations of the six-body molecular clusters (or molecules, 
for short) such as $p (d t \mu) e_2, d (d t \mu) e_2$ and $t (d t \mu) e_2$, e.g., for the $t \mu + {\rm D}_2$ 
reaction we can write  
\begin{eqnarray}
 t \mu + {\rm D}_2 &=& \Bigl\{[d (d t \mu)] e_2\Bigr\}^{*}  \; \; , \; \label{res}
\end{eqnarray}
where the notation $*$ means internal excitation of the molecular $[d (d t \mu)] e_2$ system, i.e., rotational 
and/or vibrational excitations of this two-center, six-body and two-electron molecule. In this case the rate 
of formation of the three-body muonic $d d \mu$ and $d t \mu$ ions increases substantially and it approximately 
equals to the frequency of inter-molecular transitions between different vibrational and rotational levels in 
the usual two-center molecules, e.g., in the D$_2$ and/or DT molecules. The last value can be evaluated as $3 
\cdot 10^{12} - 0.5 \cdot 10^{12}$ $sec^{-1}$. This means that formation of the three-body $d d \mu$ and $d t 
\mu$ ions takes a very short time $\approx 1 \cdot 10^{-12}$ $sec$. Certainly, for resonance muon-catalyzed 
nuclear fusion the formation of the three-body $d d \mu$ and $d t \mu$ ions is not a limiting step. 

Almost immediately after the paper \cite{Bel} was published (in fact, already in 1960) an intense stream of 
speculations has started that these weakly-bound (1,1)-states could help to increase the fusion coefficients 
$\kappa_{dd}$ and $\kappa_{dt}$ to very large values, if we consider the process in liquid deuterium and 
deuterium-tritium mixtures, respectively. A few years later a number of experiments have been performed 
\cite{Fus1}, \cite{Fus2} (see also \cite{Fus3} and references therein). It was shown that in liquid deuterium 
the corresponding $\kappa$ numbers for the $(d,d)-$ and $(d,t)-$ fusion reactions are around $\kappa_{dd} 
\approx$ 15 - 20 and $\kappa_{dt} \approx$ 90 - 120, respectively. Twenty years later these values have been 
increased up to $\kappa_{dd} \approx$ 30 for the $d d \mu$ ions and $\kappa_{dt} \approx$ 155 - 185 for the 
$d t \mu$ ions (see, e.g., \cite{Balin}, \cite{Breun} and references therein). All references to experimental 
papers, which have been performed to investigate the resonance nuclear fusion in the $d d \mu$ and $d t \mu$ 
ions, prior 1990 can be found in \cite{Petr} (see also \cite{MS}). 

In order to evaluate a possibility to apply the muon-catalysed fusion for energy production purposes let us 
assume that one negatively charged muon can catalyze exactly 200 $(d,t)-$fusion reactions. Currently, such 
a number of fusion reactions per one muon seems to be large, but it close to the upper limit and we just 
want to evaluate how much $(d,t)-$fusion reactions is sufficient to reach break-even. As is well known the 
thermal energy released during one $(d,t)-$reaction is $\approx$ 3.5 $MeV$, while contribution from the final 
14.1 $MeV$ neutron is always ignored, since there is no way to utilize this energy into something useful at 
regular densities, temperatures and pressures. On the other hand to reach theoretical break-even we need to 
compensate for the energy spent for creation of one negatively charged muon $\mu^{-}$ ($\approx$ 8000 $MeV$ 
= 0.8 $GeV$). Thus, to reach the break-even one negatively charged muon must catalyze $\kappa \approx$ 2,285 
nuclear $(d,t)$ reactions. This evaluation is still far to optimistic, since at real conditions only 70\% of 
all muons can catalyze the maximal number of fusion reactions during their life-time $\tau_{\mu} \approx 2.1 
\cdot 10^{-6}$ $sec$. Furthermore, any thermal-to-electricity conversion has less than $\approx$ 30 \% 
efficiency. From here one finds that the factor $\kappa$ must be around 8,500 - 11,000, i.e., in 40 - 55 
times larger than it is now. Such a huge deviation explains why the muon-catalysed nuclear fusion will never 
be used for the energy productions. 

Note also that the muon-catalyzed fusion proceeds (with relatively high efficiency) only in the liquid 1:1 
deuterium-tritium mixture with boiling temperature $\approx$ - 253${}^{\circ}C$. Any substantial energy 
release will produce intense heating and evaporation of tritium which is very dangerous for working 
personnel and surrounding areas. Furthermore, in the case of substantial overheating our `ideal' conditions 
for the resonance muon-catalyzed fusion will suddenly be lost. In other words, the crucial factor $\kappa$ 
will decrease instantly and substantially. Therefore, in respect to the Carnot's theorem the overall 
efficiency $\eta = \frac{T_{max} - T_{min}}{T_{max}}$ of any device based on some thermal `resonance' 
process, including the resonance muon-catalysed fusion, can never be high. Indeed, for similar `resonance' 
thermal systems the maximal temperature $T_{max}$ cannot deviate substantially from the minimal temperature 
$T_{min}$, since otherwise the optimal conditions for the resonance process will be lost.  
 
\section{Universal variational expansion for Coulomb three-body systems}
\label{B} 

Our variational expansions, Eqs.(\ref{exprel}) and (\ref{expu}), provide a number of advantages for bound 
state, highly accurate computations of different three-body systems. Applicability of these variational 
expansions to various one-center (or atomic-like) three-body systems, muonic molecular ions and exotic 
three-body systems with three comparable masses of particles, etc, has been illustrated in numerous papers 
and review articles. So, it was a great temptation to declare and consider our variational expansions, 
Eqs.(\ref{exprel}) and (\ref{expu}), as the universal three-body expansions which provide high and very 
high accuracy for arbitrary three-body systems. However, in 1987 we have found \cite{Fro87} that these 
variational expansions rapidly lose their overall accuracy, if they are applied to the two-center Coulomb 
three-body systems such as H$^{+}_{2}$, HD$^{+}_{2}$, T$^{+}_{2}$ ions and other similar molecular systems. 
For heavier two-center systems, the overall accuracy of the results obtained with the use of variational 
expansions, Eqs.(\ref{exprel}) and (\ref{expu}), rapidly decreases when the masses of heavy particles 
increase. In general, the class of two-center (or adiabatic) molecular systems is very wide and includes a 
large number of actual three-body systems which are of great interest in the molecular and plasma physics, 
astrophysics, solid state physics, etc. Therefore, we cannot simply ignore such a wide class of three-body 
systems/ions and have to modify our variational expansions in the relative and/or perimetric coordinates 
in order to achieve very high accuracy even for such systems. Our current variational expansions, 
Eqs.(\ref{exprel}) and (\ref{expu}), can be recognized as highly accurate for all three-body systems, 
except only the two-center three-body systems/ions $a^{+} a^{+} e^{-}$ (= $a a e$) and $a^{+} b^{+} e^{-}$ 
(= $a b e$), where the mass ratio $\tau = \frac{M_{min}}{m_{min}} = \frac{M_{min}}{m_e}$, where $M_{min} = 
\min (M_a, M_b)$, is large and very large, e.g., $\tau \ge 1000 - 1500$.   

Let us discuss here the current situation with highly accurate, bound state computations of the two-center, 
three-body $a a e$ and $a b e$ ions, which have larger mass ratios, e.g., for the systems with $\tau \ge$ 
1000 - 1500. Similar problem has been considered in \cite{Sutclf}. As mentioned above, the convergence rate 
of our variational expansions Eqs.(\ref{exprel}) and (\ref{expu}) rapidly decreases when the mass ratio 
$\tau$ increases. In our first calculations of the ${}^{1}$H$^{+}_{2}$ molecular ion it was hard to obtain 
even four correct decimal digits. For the model, truly adiabatic ${}^{\infty}$H$^{+}_{2}$ molecular ion 
analogous results were much worse. It was clear that the both variational expansions, Eqs.(\ref{exprel}) 
and (\ref{expu}), must somehow be modified and substantially upgraded. Our work in this direction started 
in early 1990's and finished almost ten years later, when the final paper \cite{Fro2002} was published. In 
\cite{Fro2002} we have shown that our variational expansions, Eqs.(\ref{exprel}) and (\ref{expu}), can 
easily be transformed into highly accurate, variational expansions, if all non-linear parameters 
$\alpha_{i}, \beta_{i}, \gamma_{i}$, where $i = 1, \ldots, N$, in these expansions are chosen (and then 
optimized) as complex numbers. This means that our exponential variational expansions in the relative and 
perimetric coordinates can be written for $L = 0$ in one of the two following forms  
\begin{eqnarray}
  & & \Psi(r_{32}, r_{31}, r_{21}) = \sum_{i=1}^{N} C_{i} \exp(-\alpha_{i} r_{32} - \beta_{i} r_{31} - 
 \gamma_{i} r_{21} - \imath \delta_{i} r_{32} - \imath e_{i} r_{31} - \imath f_{i} r_{21}) \; \; 
 \label{complexp0} 
\end{eqnarray}
and
\begin{eqnarray}
  \Psi(u_1, u_2, u_3) = \sum_{i=1}^{N} C_{i} \exp(-\alpha_{i} u_1 - \beta_{i} u_2 - \gamma_{i} u_3 - 
  \imath \delta_{i} u_1 - \imath e_{i} u_2 - \imath f_{i} u_3) \; , \; \label{complexp} 
\end{eqnarray}
where all $6 N$ non-linear parameters $\alpha_{i}, \beta_{i}, \gamma_{i}, \delta_i, e_{i}$ and $f_{i}$ 
($i$ = 1, $\ldots, N$) are the real numbers. Recently, the exponential expansion, Eq.(\ref{complexp}), 
has successfully been applied (see, e.g., \cite{Fro2018A}, \cite{Fro2019A} and \cite{Fro2021})to various 
three-body systems, including Coulomb three-body systems and adiabatic (or two-center) molecular ions 
such as $a^{+} a^{+} e^{-}$ (or $a a e$) and $a^{+} b^{+} e^{-}$ (or $a b e$), where $M_b \ge M_a \ge$ 
25,000 $m_e (= 1)$. For instance, in \cite{{Fro2021}}, \cite{Fro2018A} and \cite{Fro2021} we have shown 
that this variational expansion is highly accurate not only for the $p p e, d d e, t t e$ ions, but also 
for other similar ions in which $min (M_a, M_b) \le$ 50,000 - 100,000 $m_e$. Even for the two-center 
molecular ions in which $min (M_a, M_b) \le$ 300,000 - 350,000 $m_e$ ($\tau \ge$ 350,000 $m_e$) our 
variational expansion provides very good numerical accuracy ($\approx 1 \cdot 10^{-12}$ $a.u.$ - 1 $\cdot 
10^{-11}$ $a.u.$) in the bound state calculations of the $a^{+} b^{+} e^{-}$ ions. Furthermore, by 
optimizing (very carefully) all internal non-linear parameters of our method such a `boundary' mass ratio 
can be increased up to 500,000 $m_e$. 

Note that in reality, it is hard to imagine a point particle which has the unit electrical charge and mass 
$\ge$ 25,000 $m_e$. Therefore, the area of applicability of our variational expansion, Eq.(\ref{complexp}), 
already includes all Coulomb three-body systems with unit charges which do exist in the Nature. This means 
that the exponential expansion, Eq.(\ref{complexp}), already is the universal variational expansion for all 
three-body systems, including the two-center (but one-electron!) systems, which exist in Nature and can be 
studied in actual experiments. The same conclusion is true for the variational expansion, 
Eq.(\ref{complexp0}), written in the relative coordinates. However, there is some, pure academic interest 
to study the bound states in the model three-body ions with very large mass ratios $\tau$, e.g., $\tau \ge$ 
1,000,000 $m_e$. 

By analysing the bound state calculations of similar systems with very large nuclear masses we have noticed 
that nobody has ever investigated the general situation, which can be found in the model, very heavy 
two-center $a^{+} a^{+} e^{-}$ and $a^{+} b^{+} e^{-}$ ions, where $min (M_a, M_b) \ge$ 1,000,000 
$m_e$. Finally, we have decided to perform an extensive numerical investigation of bound states in the 
two-center three-body molecular ions with very heavy nuclear masses. As mentioned above for such three-body 
systems our variational expansion, Eq.(\ref{complexp}), rapidly loses its overall numerical accuracy and in 
applications to the truly adiabatic, two-center molecular ions such as ${}^{M}$H$^{+}_{2}$, where $M \ge$ 
1,000,000 $m_e$, our variational expansion, Eq.(\ref{complexp}), is not highly accurate. Formally, in this 
mass-area one finds only a few model, truly adiabatic ${}^{M}$H$^{+}_{2}$ and ${}^{\infty}$H$^{+}_{2}$ ion 
which are of a restricted, pure theoretical interest. For instance, the total energies of similar three-body 
ions are used to construct some highly accurate mass-interpolation formulas (see, e.g., \cite{Fro2021}, 
\cite{Fro2018A} and \cite{Fro2019A}) for the Coulomb three-body systems. Nevertheless, we have investigated 
this problem and found its solution, which can be represented in the following form. In order to perform 
highly accurate, bound state calculations of the truly adiabatic, two-center three-body systems, including 
the model ${}^{\infty}$H$^{+}_{2}$ ion and close systems our variational expansion, Eq.(\ref{complexp}), 
must be modified by adding one additional Gauss-like factor in each basis function. Probably, this is the 
shortest way to reach our goals and modify our variational expansion, Eq.(\ref{complexp}) for the truly 
adiabatic, two-center (three-body) systems. 

This leads to the following explicit form of this `upgraded', highly accurate, variational expansion for 
the bound states with $L = 0$ in the relative and/or perimetric coordinates:
\begin{eqnarray}
  \Psi(r_{32}, r_{31}, r_{21}) &=& \sum_{i=1}^{N} C_{i} \exp\Bigl[-\alpha_{i} r_{32} - \beta_{i} r_{31} 
   - \gamma_{i} r_{21} \Bigr] \; \exp\Bigl[ - f_i r^{2}_{21} \Bigr] \; \; \nonumber \\
  &=& \sum_{i=1}^{N} C_{i} \exp\Bigl[-\alpha_{i} r_{32} - \beta_{i} r_{31} - \gamma_{i} r_{21} - f_i 
       r^{2}_{21} \Bigr] \; \; , \; \; \label{vibr} 
\end{eqnarray}
where the symbols 1 and 2 designate the very heavy (even infinitely heavy) atomic nuclei, while the 
particle 3 denotes the electron $e^{-}$. The parameters $\alpha_i, \beta_i, \gamma_i$ ($i = 1, 2, \ldots, 
N$) in Eq.(\ref{vibr}) are the complex numbers which are optimized, while the parameters $f_i$ ($i = 1, 
2, \ldots, N$) are the real positive numbers. The variational expansion, Eq.(\ref{vibr}), can also be 
written in a number of different forms. For instance, one of the first version of our variational 
expansion of this kind was quite similar to Eq.(\ref{complexp}), but each wave function included the 
same Gauss-like factor with one additional, $i-$independent non-linear parameter $f (= f_i)$. Note that 
all these variational expansions developed for highly accurate computations of pure adiabatic three-body 
ions and similar model systems must include the Gauss-like factor upon the inter-nuclear $r_{21}$ variable. 
This factor is crucially important to obtain results (energies) of high numerical accuracy for the truly 
adiabatic systems, including the model ${}^{\infty}$H$^{+}_{2}$ ion. The physical sense of this factor in 
each basis function seems to be obvious: it is explicitly describes slow nuclear vibrations in the very 
heavy and truly adiabatic ${}^{M}$H$^{+}_{2}$ and ${}^{\infty}$H$^{+}_{2}$ ions, where $M \ge \;$ 
1,000,000 $m_e$. In general, an accurate description of similar low frequency oscillations (or vibrations) 
in the molecular two-center ions by our exponents from Eqs.(\ref{complexp0}) and (\ref{complexp}) is a 
difficult problem. In other words, the variational Fourier-like expansions of trial wave functions have 
very slow convergence rate(s) for the truly adiabatic systems. An additional Gauss-like factor in 
Eq.(\ref{vibr}) allows one to speed up the overall convergence rate qualitatively, e.g., in thousands and 
even millions times. In applications to the truly adiabatic systems with very large nuclear masses, e.g., 
$\Bigl(\frac{M}{m_e}\Bigr) \ge 1 \cdot 10^{7}$, such an acceleration is critical. For the first time the 
variational expansion, Eq.(\ref{vibr}), was proposed and used by N.N. Kolesnikov and his co-workers (see, 
e.g., \cite{Kol1984}, \cite{Kol2004} and earlier references therein). However, at that time nobody 
considered this expansion as some universal variational expansion for three-body systems.     

The new variational expansion, Eq.(\ref{vibr}), includes $4 N$ non-linear parameters which provide a very 
good flexibility of the constructed (trial) wave functions. We hope to describe applications of this new 
variational, universal three-body expansion, Eq.(\ref{vibr}), in our next article. Here we just want to 
present only some preliminary numerical results. First of all, we have to note that the new variational 
expansion, Eq.(\ref{vibr}), has a very high convergence rate in bound state computations of the truly 
adiabatic ions. For instance, by using 75, 200 and 300 basis functions in Eq.(\ref{vibr}) we have found 
the following total energies of the ground $1 s \sigma$ state in the truly adiabatic, model 
${}^{\infty}$H$^{+}_{2}$ ion: $E(75)$ = -0.60263421 025 $a.u.$, $E(275)$ = -0.60263421 44948595 $a.u.$ 
and $E(500)$ = -0.60263421 44949462 5807 $a.u.$ The current value of the exact total energy of this 
bound $1 s \sigma$ state is $\approx$ -0.60263421 44949464 6150905(5) $a.u.$ As follows from these 
results \cite{H2+New} we do observe a very fast convergence of the computed energies to the exact value. 
However, we have to note that all bound state calculations of the model ${}^{\infty}$H$^{+}_{2}$ ion with 
the use of our new universal three-body, variational expansion, Eq.(\ref{vibr}), are quite difficult to 
perform. The main reason is quite obvious, since the overlap matrices for all relatively large dimensions 
rapidly become ill-conditioned. For instance, we have found that for $N \ge 350 - 375$ the overlap matrices 
are already extremely ill-conditioned. Therefore, it is absolutely necessary to apply some special codes 
and packages for numerical calculations with arbitrary precision (see, e.g., \cite{Bail1} and \cite{Bail2}) 
and constantly increase (when $N$ increases) the number of decimal digits per one computer word in similar 
calculations.      
  
On the other hand, we have to remember that the truly adiabatic and model ${}^{\infty}$H$^{+}_{2}$ ion is 
a one-electron, two-center system. There is an additional symmetry in similar systems, since the Coulson's 
operator $\hat{\Lambda} = \hat{L}^{2} + \sqrt{- 2 E} \; R \; \hat{A}_z$ is rigorously conserved in the 
pure-adiabatic, two-center ${}^{\infty}$H$^{+}_{2}$ ion and approximately conserved in close systems. Here 
$\hat{L}^{2}$ and $\hat{A}_z$ are the operators of angular (electronic) moment and $z-$component of the 
Runge-Lenz vector-operator  $\hat{\bf A}$ \cite{Runge}, \cite{Lenz}, \cite{Lapl}. Also, in this formula 
$R$ is the inter-nuclear distance and $E \le 0$ is the total energy of the bound state in this three-body 
ion (for more details, see, e.g., \cite{Fro83} and references therein). The operator $\hat{\Lambda}$ is 
responsible for exact separation of radial and angular spheroidal variables (or coordinates) and this 
allows one to reduce the original two-dimensional problem to the solution of two one-dimensional 
differential equations. As follows from these arguments the original problem is much simpler than our new 
advanced (but three-body!) methods which are currently applied to solve it. Note also that a large number 
of very effective semi-analytical and pure numerical methods have been developed to solve many problems 
for one-electron, truly adiabatic (or two-center) molecular ions since 1940's (see, e.g., \cite{MQS}, 
\cite{Eyr1}, \cite{Coul}, \cite{Slav} and references therein). A number of properties of the model 
${}^{\infty}$H$^{+}_{2}$ ion in its ground (bound) $1 s \sigma$-state were accurately evaluated in these 
earlier papers. 

Currently, we can substantially improve the overall accuracy and efficiency of these `old' methods, by 
performing computations based on arithmetic operations with arbitrary precision, flexible memory relocation 
and other modern tricks. For instance, in 1970's all calculations of the ground (bound) $1 s \sigma-$state 
in the model ${}^{\infty}$H$^{+}_{2}$ ion have been performed with the use of 5,000 - 10,000 very simple 
basis functions (examples can be found in \cite{Slav}). It was sufficient to produce 11 - 13 exact (or 
stable) decimal digits for the total energy which was absolutely normal at that time. Nowadays, one can 
increase the numbers of basis functions used up to 100,000,000 and more (even much more). This should allow 
one to obtain 40 - 55 exact (or stable) decimal digits for the total ground energy of the truly adiabatic, 
model ${}^{\infty}$H$^{+}_{2}$ ion in its $1 s \sigma-$state. Here we cannot discuss these traditional 
methods which have substantially been improved. Now, these methods are ready for numerical applications to 
the Coulomb two-center systems known in quantum mechanics.

\newpage
\begin{table}[tbp]
   \caption{The total energies in muon-atomic units ($m.a.u.$) of weakly-bound (1,1)-states in the 
            $d d \mu$ and $d t \mu$ muonic ions and (2,0)-state in the $d t \mu$ muonic ion. The 
            notation $N$ stands for the total number of exponential basis functions used.}
     \begin{center}
    \scalebox{1.00}{%
     \begin{tabular}{| c | c | c | c |}
      \hline\hline
   $N$ & $d d \mu$ ion; $(1,1)-$state & $N$ & $d t \mu$ ion; $(1,1)-$state \\ 
     \hline
 3500 & -0.4736867 2970969 890366 & 3500 & -0.4819915 2652986 44014 \\  

 3600 & -0.4736867 2970969 890455 & 3600 & -0.4819915 2652986 45734 \\  

 3700 & -0.4736867 2970969 890539 & 3700 & -0.4819915 2652986 47858 \\

 3800 & -0.4736867 2970969 890593 & 3800 & -0.4819915 2652986 48961 \\

 3840 & -0.4736867 2970969 890621 & 3840 & -0.4819915 2652986 49445 \\ 

 3842 & -0.4736867 2970969 890623 & 3842 & -0.4819915 2652986 49455 \\     
              \hline\hline      
  $N$ & $d t \mu$ ion; $(2,0)-$state & $N$ & $d t \mu$ ion; $(2,0)-$state \\ 
     \hline
  4500 & -0.500118077322051 & 5100 & -0.500118077325339 \\

  4700 & -0.500118077322071 & 5300 & -0.500118077325419 \\

  4900 & -0.500118077323436 & 5500 & -0.500118077325475 \\
      \hline\hline  
   $N$ & $d d \mu$ ion; $(2,0)-$state & $N$ & $t t \mu$ ion; $(2,0)-$state \\ 
     \hline
  2800 & -0.4887083242297 & 3000 & -0.5125686232381 \\

  3000 & -0.4887083245714 & 3000 & -0.5125686334174 \\

  3200 & -0.4887083247575 & 3200 & -0.5125686375020 \\

  3400 & -0.4887083248687 & 3400 & -0.5125686421485 \\

  3600 & -0.4887083249605 & 3600 & -0.5125686439316 \\  

  3700 & -0.4887083249914 & 3700 & -0.5125686446090 \\

       &                  & 3800 & -0.5125686450010 \\
      \hline\hline    
  \end{tabular}}
  \end{center}
  \end{table}
\begin{table}[tbp]
   \caption{The expectation values of a number of regular properties (in muon-atomic units) of the weakly-bound 
           (1,1)-state in the muonic $d t \mu$ ion. The index 3 stands for the negatively charged $\mu^{-}$ ion, 
           the index 1 means the deuterium nucleus $d^{+}$, while the index 2 designates the tritium nucleus 
           $d^{+}$ in the muonic $d t \mu$ ion.}
     \begin{center}
     \scalebox{0.95}{%
     \begin{tabular}{| c | c | c | c |}
      \hline\hline
  $\langle \frac12 {\bf p}^{2}_{3} \rangle$ & 0.451072235514906 & 
              $\langle r^{-1}_{21} \rangle$ & 0.156988598283185 \\

  $\langle \frac12 {\bf p}^{2}_{2} \rangle$ & 0.215766575450378 & 
              $\langle r^{-1}_{31} \rangle$ & 0.301826947193663 \\

  $\langle \frac12 {\bf p}^{1}_{3} \rangle$ & 0.498854977229344 & $\langle r^{-1}_{32} \rangle$ 
                   & 0.819144704149248 \\
                   \hline
  $\langle {\bf p}_{1} \cdot {\bf p}_{2} \rangle$ & -0.131774658582408 & 
  $\langle \frac{{\bf r}_{31} \cdot {\bf r}_{32}}{r^{3}_{31}} \rangle$ & -0.400520547403092 \\

  $\langle {\bf p}_{1} \cdot {\bf p}_{3} \rangle$ & -0.083991916867970 & 
  $\langle \frac{{\bf r}_{31} \cdot {\bf r}_{21}}{r^{3}_{21}} \rangle$ & -0.32148537218059 \\

  $\langle {\bf p}_{2} \cdot {\bf p}_{3} \rangle$ & -0.36708031864694 & 
  $\langle \frac{{\bf r}_{32} \cdot {\bf r}_{12}}{r^{3}_{32}} \rangle$ & 0.03817172174731 \\
                    \hline
  $\langle r_{21} \rangle$ & 2.0648848561192 & $\langle r^{2}_{21} \rangle$ & 20.9577469505 \\

  $\langle r_{31} \rangle$ & 2.6704405487459 & $\langle r^{2}_{31} \rangle$ & 53.6374477311 \\ 

  $\langle r_{32} \rangle$ & 1.1997897390158 & $\langle r^{2}_{32} \rangle$ & 3.23366303337 \\
                   \hline
  $\langle {\bf r}_{32} \cdot {\bf r}_{31} \rangle$ & -3.79821963625 & $\langle r^{3}_{21} \rangle$ & 252.5469057 \\ 

  $\langle {\bf r}_{31} \cdot {\bf r}_{21} \rangle$ & 13.92586428087 & $\langle r^{3}_{31} \rangle$ & 98.01053721 \\

  $\langle {\bf r}_{21} \cdot {\bf r}_{32} \rangle$ &  7.03188266962 & $\langle r^{3}_{32} \rangle$ & 27.59863151 \\
    \hline\hline
  \end{tabular}}
  \end{center}
  \end{table}
%
%
%
%
\begin{table}[tbp]
   \caption{The expectation values of a number of regular properties (in muon-atomic units) of the rotationally 
            excited (2,0)-state in the muonic $d t \mu$ ion. The index 3 stands for the negatively charged 
            $\mu^{-}$ ion, the index 1 means the deuterium nucleus $d^{+}$, while the index 2 designates the 
            tritium nucleus $d^{+}$ in the muonic $d t \mu$ ion.}
     \begin{center}
     \scalebox{0.95}{%
     \begin{tabular}{| c | c | c | c |}
      \hline\hline
  $\langle \frac12 {\bf p}^{2}_{3} \rangle$ & 0.4410299320417745 & 
              $\langle r^{-1}_{21} \rangle$ & 0.3050423341412470 \\

  $\langle \frac12 {\bf p}^{2}_{2} \rangle$ & 0.6575222640866448 & 
              $\langle r^{-1}_{31} \rangle$ & 0.5590590673472879 \\

  $\langle \frac12 {\bf p}^{1}_{3} \rangle$ & 0.6098632803783084 &  
              $\langle r^{-1}_{32} \rangle$ & 0.7462194214169826 \\
                   \hline
  $\langle {\bf p}_{1} \cdot {\bf p}_{2} \rangle$ & -0.4131778062115894 & 
  $\langle \frac{{\bf r}_{31} \cdot {\bf r}_{32}}{r^{3}_{31}} \rangle$ & -0.563840009931255 \\

  $\langle {\bf p}_{1} \cdot {\bf p}_{3} \rangle$ & -0.1966854741667190 & 
  $\langle \frac{{\bf r}_{31} \cdot {\bf r}_{21}}{r^{3}_{21}} \rangle$ &  0.889634446810501 \\

  $\langle {\bf p}_{2} \cdot {\bf p}_{3} \rangle$ & -0.2443444578750554 & 
  $\langle \frac{{\bf r}_{32} \cdot {\bf r}_{12}}{r^{3}_{32}} \rangle$ & -0.077654003522546 \\
                    \hline
  $\langle r_{21} \rangle$ & 5.5779173588 & $\langle r^{2}_{21} \rangle$ & 35.3318560 \\

  $\langle r_{31} \rangle$ & 2.1117084584 & $\langle r^{2}_{31} \rangle$ & 6.44872887 \\ 

  $\langle r_{32} \rangle$ & 2.3103741828 & $\langle r^{2}_{32} \rangle$ & 7.40285538 \\
                   \hline
  $\langle {\bf r}_{32} \cdot {\bf r}_{31} \rangle$ & -10.74013589 & $\langle r^{3}_{21} \rangle$ & 62.7408 \\ 

  $\langle {\bf r}_{31} \cdot {\bf r}_{21} \rangle$ &  17.18886476 & $\langle r^{3}_{31} \rangle$ & 6.43083 \\
 
  $\langle {\bf r}_{21} \cdot {\bf r}_{32} \rangle$ &  18.14299127 & $\langle r^{3}_{32} \rangle$ & 7.68638 \\
    \hline\hline
  \end{tabular}}
  \end{center}
  \end{table}
%
%
%
%
\begin{table}[tbp]
   \caption{The expectation values of a number of regular properties (in muon-atomic units) of the rotationally excited
           (2,0)-state in the muonic $d d \mu$ and $t t \mu$ ions. The index 3 stands for the negatively charged 
           $\mu^{-}$ ion, while the indexes 1 and 2 designate the two identical nuclei of hydrogen isotopes, i.e., 
           deuterium (or tritium) in the muonic $d d \mu$ and $t t \mu$ ions.}
     \begin{center}
     \scalebox{0.95}{%
     \begin{tabular}{| c | c | c | c | c | c |}
      \hline\hline
       & $d d \mu$ & $t t \mu$ &  & $d d \mu$ & $t t \mu$ \\
          \hline
  $\langle \frac12 {\bf p}^{2}_{3} \rangle$ & 0.4252930422 & 0.4581822407 & $\langle r^{-1}_{21} \rangle$ & 
                                              1.2607674155 & 1.3549142359 \\

  $\langle \frac12 {\bf p}^{2}_{1} \rangle$ & 0.5628636532 & 0.7229287641 & $\langle r^{-1}_{31} \rangle$ & 
                                              0.9466770026 & 1.0983267005 \\
                  \hline
  $\langle {\bf p}_{1} \cdot {\bf p}_{2} \rangle$ & -0.3502171321 & -0.4938376400 & $\langle \frac{{\bf r}_{31} 
   \cdot {\bf r}_{32}}{r^{3}_{31}} \rangle$ & -0.9236353281 & -0.9982092663 \\

  $\langle {\bf p}_{1} \cdot {\bf p}_{3} \rangle$ & -0.2126465211 & -0.2290911203 & 
  $\langle \frac{{\bf r}_{31} \cdot {\bf r}_{21}}{r^{3}_{21}} \rangle$ & 1.6367022405 & 1.7849696206 \\
                    \hline
  $\langle r_{21} \rangle$ & 5.525945087 & 5.626231716 & $\langle r^{2}_{21} \rangle$ & 34.6982928 & 35.8976207 \\

  $\langle r_{31} \rangle$ & 2.237027453 & 2.186884133 & $\langle r^{2}_{31} \rangle$ & 7.06137966 & 6.77317592 \\ 
                   \hline
  $\langle {\bf r}_{32} \cdot {\bf r}_{31} \rangle$ & -10.28776671 & -11.17563442 & $\langle r^{3}_{21} \rangle$ & 
                                                      62.43918 & 62.89322 \\ 

  $\langle {\bf r}_{31} \cdot {\bf r}_{21} \rangle$ & 17.34914637 & 17.94881034 & $\langle r^{3}_{31} \rangle$ & 
                                                      8.137607 & 9.363156 \\
    \hline\hline
  \end{tabular}}
  \end{center}
  \end{table}
%
%
%
%
\begin{table}[tbp]
   \caption{The expectation values of a number of regular properties (in muon-atomic units) of the weakly-bound 
           (1,1)-state in the muonic $d d \mu$ ion. The index 3 stands for the negatively charged $\mu^{-}$ ion, 
           the index $o$ means the outermost deuterium nucleus, while the index $c$ designates the closest 
           deuterium nucleus $d^{+}$ in the muonic $d d \mu$ ion.}
     \begin{center}
     \scalebox{0.95}{%
     \begin{tabular}{| c | c | c | c |}
      \hline\hline
  $\langle \frac12 {\bf p}^{2}_{\mu} \rangle$ & 0.437472785565039 & 
                $\langle r^{-1}_{dd} \rangle$ & 0.146140889429525 \\

  $\langle \frac12 {\bf p}^{2}_{d} \rangle$ &    0.32142904911275 & 
            $\langle r^{-1}_{d\mu} \rangle$ &    0.54675717442446 \\
                   \hline
  $\langle {\bf p}_{1} \cdot {\bf p}_{2} \rangle$ & -0.1026926563302 & 
  $\langle \frac{{\bf r}_{31} \cdot {\bf r}_{32}}{r^{3}_{31}} \rangle$ & -0.4005205474031 \\

  $\langle {\bf p}_{1} \cdot {\bf p}_{3} \rangle$ & -0.2187363927825 & 
  $\langle \frac{{\bf r}_{31} \cdot {\bf r}_{21}}{r^{3}_{21}} \rangle$ & -0.32148537218058 \\
                    \hline
  $\langle r_{dd} \rangle$   & 2.0413404513094 & $\langle r^{2}_{dd} \rangle$   & 20.4685047204 \\

  $\langle r_{o\mu} \rangle$ & 2.8548174537107 & $\langle r^{2}_{o\mu} \rangle$ & 11.2793472283 \\ 

  $\langle r_{c\mu} \rangle$ & 1.0625016436705 & $\langle r^{2}_{c\mu} \rangle$ & 2.74750509179 \\
                   \hline
  $\langle {\bf r}_{32} \cdot {\bf r}_{31} \rangle$ & -3.22082620018 & $\langle r^{3}_{dd} \rangle$   & 242.5185809 \\ 

  $\langle {\bf r}_{31} \cdot {\bf r}_{21} \rangle$ &  5.96833129197 & $\langle r^{3}_{o\mu} \rangle$ & 28.14969137 \\

  $\langle {\bf r}_{21} \cdot {\bf r}_{32} \rangle$ &  14.5001734285 & $\langle r^{3}_{c\mu} \rangle$ & 11.24778142 \\
    \hline\hline
  \end{tabular}}
  \end{center}
  \end{table}
%
%
%
%
%

\begin{thebibliography}{99}

\bibitem{FroW2011} A.M. Frolov and D.M. Wardlaw, Europ. Phys. Journ. D {\bf 63} 339 (2011); ibid, {\bf 66} 
212 (2012).

\bibitem{Fro1992} A.M. Frolov, Jour. Phys. B {\bf 25}, 3059 (1992). 

\bibitem{FroPLA} A.M. Frolov, Phys. Lett. A {\bf 389}, 1288 (2019).

\bibitem{MS} L.I. Menshikov and L.N. Somov, Usp. Fiz. Nauk {\bf 160} 47, (1990) [Sov. Phys. Usp. {\bf 33}, 616 (1990)]. 
 
\bibitem{Alv} L.W. Alvarez, H. Bradner, F.S. Crawford, Jr., J.A. Crawford, P. Falk-Vairant, M.L. Good, 
J.D. Gow, A.H. Rosenfeld, F. Solmitz, M.L. Stevenson, H.K. Ticho, and R.D. Tripp, Phys. Rev. {\bf 105}, 
1127 (1957).

\bibitem{Fro56body} A.M. Frolov, European Phys. Jour. D {\bf 72}, 222 (2018). 

\bibitem{Fro2001} A.M. Frolov, Phys. Rev. E {\bf 64}, 036704 (2001); ibid, {\bf 74} 027702 (2006). 

\bibitem{Varsh} D.A. Varshalovich, A.N. Moskalev and V.K. Khersonskii, \textit{Quantum Theory of Angular 
Momentum}, (Nauka, Moscow (1975)).

\bibitem{Pek1} C.L. Pekeris, Phys. Rev. {\bf 115} 1216 (1959); ibid, {\bf 112}, 1649 (1958). 

\bibitem{Fro2021} A.M. Frolov, Jour. Mol. Phys. {\bf 119}, e1837973 (2021). 

\bibitem{MQS} R. McWeeny and B.T. Sutcliffe, \textit{Methods of Molecular Quantum Mechanics} (Acad. Press, 
New York (1969)), Chpt. 7.  

\bibitem{XXX} For the four-body systems one can also introduce six perimetric coordinates (see, e.g., 
A.M. Frolov, Jour. Phys. A {\bf 64} 036704 (2007)). Unfortunately, these coordinates do not have the same 
properties as the three-body perimetric coordinates and this fact drastically complicates applications 
of the perimetric coordinates for highly accurate, bound state computations of four-body systems.  

\bibitem{Rudin} W. Rudin, \textit{Principles of Mathematical Analysis}, (2nd ed., McGraw Hill Book Comp., New York,
(1964)).

\bibitem{Rose} M.E. Rose, \textit{Elementary Theory of Angular Momentum}, (Dover Publ. Inc., Mineola, N.Y. 
(1995)).

\bibitem{Edm} A.R. Edmonds, \textit{Angular Momentum in Quantum Mechanics}, (Princeton Univ. Press., 
Princeton, NJ (1974)). 

\bibitem{Sob} I.I. Sobel'man {\it An Introduction to the Theory of Atomic Spectra}. (Pergamon Press, N.Y. (2014)), 
Chapt. IV. 

\bibitem{FroF} C. Froese Fischer, T. Brage, P. J\"{o}nsson, \textit{Computational Atomic Structure. An MCHF Approach} 
(IOP Publishing, Bristol (1997)), Appendix A. 

\bibitem{FroSm96} A.M. Frolov and V.H. Smith, Jr., Phys. Rev. A {\bf 53}, 3853 (1996).

\bibitem{Fro87} A.M. Frolov, Zh. Eksp. Teor. Fiz. {\bf 92}, 1959 (1987) [Sov. Phys. JETP {\bf 92}, 1100 
(1987)].

\bibitem{GR} I.S. Gradstein and I.M. Ryzhik, \textit{Tables of Integrals, Series and Products} 
(6th revised ed., Academic Press, New York, (2000)).

\bibitem{NIST} see, e.g., https://physics.nist.gov/cgi-bin/cuu/Value?

\bibitem{Vinn} S.I. Vinnitskii, V.S. Melezhik, L.I. Ponomarev, I.V. Puzynin, T.P. Puzynina, 
L.N. Somov and N.F. Truskova, Zh. Eksp. Teor. Fiz. \textbf{79}, 698 (1980) [Sov. Phys. JETP 
\textbf{52}, 353 (1980)]. 

\bibitem{BD1} A.K. Bhatia and R.J. Drachman, Phys. Rev. A \textbf{30}, 2138 (1984).

\bibitem{Kam} M. Kamimura, Phys. Rev. A \textbf{38}, 621 (1988). 

\bibitem{YYY} Only for the true adiabatic and close three-body ions, e.g., for the ${}^{\infty}$H$^{+}_{2}$ 
ion, the ratio $\tau$ increases to 0.17 (or 17\%).   

\bibitem{Dirac} P.A.M. Dirac, \textit{The Principles of Quantum Mechanics} (4th ed., Oxford at the Clarendon 
Press, Oxford, UK (1958)). 

\bibitem{Migd} A. B. Migdal and V. Krainov, \textit{Approximation Methods in Quantum Mechanics} (W. A. Benjamin, 
Inc., New York (1969)).


\bibitem{Frank} F.C. Frank, Nature \textbf{160}, 525 (1947). 

\bibitem{Bel} V.B. Belyaev, S.S. Gershtein, B.N. Zahar'ev, S.P. Lomnev, Zh. Eksp. Teor. Fiz. {\bf 37}, 1652 (1959).

\bibitem{Fus1} V.P. Dzelepov, P.F. Ermolov, Yu.V. Katashev, V.I. Moskalev, V.V. Filchenkov and M. Friml, 
Zh. Eksp. Teor. Fiz. {\bf 46}, 2042 (1964).

\bibitem{Fus2} V.P. Dzelepov, P.F. Ermolov, V.I. Moskalev and V.V. Filchenkov, Zh. Eksp. Teor. Fiz. {\bf 50}, 1235 
(1966).

\bibitem{Fus3} V.M. Bystritsky, V.P. Dzelepov, V.I. Petruchin, A.I. Rudenko, L.N. Somov, V.M. Suvorov, V.V. 
Filchenkov, G. Hemnitz, N.M. Khovansky, B.A. Kohmenko and D. Horvath, Zh. Eksp. Teor. Fiz. {\bf 76}, 460 (1979).

\bibitem{Balin} D.V. Balin, E.M. Maev, V.I. Medvedev, G.G. Semenchuk, Yu.N. Smirenin, A.A. Vorobyov, An.A. Vorobyov 
and Yu.K. Zalite, Phys. Lett. B {\bf 141}, 173 (1984). 

\bibitem{Breun} W.H. Breunlich, M. Carnelli, P. Kammel, J. Marton, P. Pawlek, J. Werner, J. Emeskal, K.M. Crowe, J. 
Kurk, A. Janett, C. Petitjean, R.H. Scherman, H. Bassy and W. Nauman, Phys. Rev. Lett. {\bf 53}, 12 (1984).

\bibitem{Petr} S.S. Gerstein, Yu.V. Petrov and L.I. Ponomarev, Usp. Fiz. Nauk {\bf 160}, 3 (1990).


\bibitem{Sutclf} B.T. Sutcliffe, Theor. Chem. Acc. \textbf{118}, 563 (2007).

\bibitem{Fro2002} A.M. Frolov, J. Phys. B {\bf 35}, L331 (2002).

\bibitem{Fro2018A} A.M. Frolov, Europ. Phys. Journal D {\bf 72}, 52 (2018).

\bibitem{Fro2019A} A.M. Frolov, Chem. Phys. Lett. {\bf 723}, 160 (2019). 

\bibitem{Kol1984} P.P. Zakharov, N.N. Kolesnikov and V.I. Tarasov, Vestnik Mosk. Univ., Ser. Fiz., Astron. {\bf 24}, 
34 (1983).

\bibitem{Kol2004} A. G. Donchev, S. A. Kalachev, N. N. Kolesnikov and V. I. Tarasov, Phys. Rev. A {\bf 69}, 034501 
(2004).

\bibitem{H2+New} For the exponential variational expansion, Eq.(\ref{complexp}), our best-to-date variational energy 
of the model ${}^{\infty}$H$^{+}_{2}$ ion is -0.60262756318795 $a.u.$ 
 
\bibitem{Bail1} D.H. Bailey, AMC Trans. Math. Soft. {\bf 21}, 379 (1995).

\bibitem{Bail2} D.H. Bailey, Comput. Sci. Eng. {\bf 2}, 24 (2000). 

\bibitem{Fro83} A.M. Frolov, JETP Letters {\bf 38}, 607 (1983). 

\bibitem{Runge} C. Runge, \textit{Vektoranalysis} (Methuen \& Co., London, UK (1923)).

\bibitem{Lenz} W. Lenz, Zeit. f. Phys. {\bf 24}, 197 (1924). 

\bibitem{Lapl} The conservation of this vector in celestial mechanics was well known to P.-S. Laplace 
in 1790's. (see, e.g., the first volume of his famous book ``Trait\'{e} \; de \; m\'{e}canique \; 
c\'{e}leste'' published by Duprat, in Paris in 1799). 

\bibitem{Eyr1} H. Eyring, J. Walter and G.E. Kimbal, \textit{Quantum Chemstry} (John Wiley \& Sons (1966)).

\bibitem{Coul} C.A. Coulson, \textit{Valence} (Oxford at the Clarendon Press, Oxford, UK (1956)), Chpt. IV.

\bibitem{Slav} V.I. Komarov, L.I. Ponomarev and S.Yu. Slavyanov, \textit{Spheroidal and Coulomb Spheroidal Functions}, 
(Nauka, Moscow (1976)). 

\end{thebibliography}
\end{document}